\documentclass{article} %
\usepackage[accepted]{icml2025}
\usepackage{times}

\usepackage{amsmath,amsfonts,bm}

\def\eqref#1{equation~\ref{#1}}

\def\1{\bm{1}}

\DeclareMathAlphabet{\mathsfit}{\encodingdefault}{\sfdefault}{m}{sl}
\SetMathAlphabet{\mathsfit}{bold}{\encodingdefault}{\sfdefault}{bx}{n}

\usepackage{adjustbox}
\usepackage{hyperref}
\usepackage{url}
\usepackage{wrapfig}
\usepackage[utf8]{inputenc} %
\usepackage[T1]{fontenc}    %
\usepackage{hyperref}       %
\usepackage{url}            %
\usepackage{booktabs}       %
\usepackage{nicefrac}       %
\usepackage{microtype}      %
\usepackage{graphicx}
\usepackage{xcolor}
\usepackage{array}
\usepackage{nth}
\usepackage{pifont}%

\usepackage{subfigure}
\usepackage{amsmath}
\usepackage{amssymb}
\usepackage{mathtools}
\usepackage{amsthm}
\usepackage{amsfonts}       %
\usepackage{multirow}
\usepackage{float}
\usepackage{cleveref}       %
\usepackage{pdfpages}
\usepackage{graphicx}
\usepackage{makecell}
\usepackage{soul}
\usepackage{enumitem}
\usepackage{algorithm}      %
\usepackage{algorithmic}
\usepackage{adjustbox}
\usepackage{comment}
\usepackage{setspace}

\newcommand{\maintitle}{%
    Hessian-aware Training for Enhancing DNNs Resilience to\\ Parameter Corruptions}
\newcommand{\runningtitle}{%
    Hessian-aware Training for Enhancing DNNs Resilience to Parameter Corruptions}

\newcommand{\topic}[1]{\noindent\textbf{#1}}

\icmltitlerunning{\runningtitle{}}

\begin{document}

\twocolumn[
\icmltitle{\maintitle{}}

\icmlsetsymbol{equal}{*}

\begin{icmlauthorlist}
\icmlauthor{Tahmid Hasan Pranto}{osu}
\icmlauthor{Seijoon Kim}{ss}
\icmlauthor{Lizhong Chen}{osu}
\icmlauthor{Sanghyun Hong}{osu}
\end{icmlauthorlist}

\icmlaffiliation{osu}{%
    Department of Computer Science, 
    Oregon State University, Corvallis, OR USA}
\icmlaffiliation{ss}{%
    Samsung Advanced Institute of Technology, Suwon, South Korea}

\icmlcorrespondingauthor{Sanghyun Hong}{sanghyun.hong@oregonstate.edu}

\icmlkeywords{Machine Learning, ICML}

\vskip 0.3in

]

\begin{abstract}
    Deep neural networks are not resilient to parameter corruptions: 
    even a single-bitwise error in their parameters in memory can 
    cause an accuracy drop of over 10\%, and in the worst-cases, up to 99\%.
    This susceptibility poses great challenges in deploying models on
    computing platforms, where adversaries can induce bit-flips 
    through software or bitwise corruptions may occur naturally.
    Most prior work addresses this issue with hardware or system-level approaches,
    such as integrating additional hardware components
    to verify a model's integrity at inference.
    However, these methods have not been widely deployed 
    as they require infrastructure or platform-wide modifications.
    
    In this paper, we propose a new approach to addressing this issue:
    training models to be more resilient to bitwise corruptions to their parameters.
    Our approach, Hessian-aware training,
    promotes models to have \emph{flatter} loss surfaces.
    We show that,
    while there have been training methods,
    designed to improve generalization through Hessian-based approaches,
    they do not enhance resilience to parameter corruptions.
    In contrast, 
    models trained with our method demonstrate increased resilience to parameter corruptions, 
    particularly with a 20--50\% reduction in
    the number of bits whose individual flipping leads to a 90--100\% accuracy drop.
    Moreover, we show the synergy between ours 
    and existing hardware and system-level defenses.
\end{abstract}

\section{Introduction}
\label{sec:intro}

Deep neural networks (DNNs) are \emph{not} resilient to parameter corruptions.
Prior work has shown that adversaries,
particularly who are capable of causing targeted bitwise errors 
on the memory representation of their parameters,
can exploit this property to induce undesirable behaviors.
This includes substantial accuracy drop%
~\cite{hong2019terminal, rakin2019bit, deephammer}, 
targeted misclassification~\cite{bai2023versatile, cai2021seeds, rakin2021t}, 
and backdoor injections~\cite{Chen_2021_ICCV, 10646710, rakin2020tbt}.
Recent work have also demonstrated privacy risks,
such as model extraction~\cite{deepsteal}.

Most prior work addresses this vulnerability
by developing defenses at the hardware-level or system-level%
~\citep{bennett2021panopticon, rakin2021ra, radar, 
        di2023copy, zhou2023dnn, zhou2023dram, neuropots, ageis}.
While having demonstrated their effectiveness,
these approaches are often difficult to implement in practice 
as they require additional hardware components 
or updates to system software, 
necessitating infrastructure-wide changes.

In this work, we study a new, \emph{orthogonal} approach 
that has not been investigated in prior work: 
enhancing a model's natural resilience to parameter corruptions.
By decreasing the number of model parameters 
whose bitwise corruptions can cause a substantial accuracy drop,
or by minimizing the performance degradation resulting from bitwise errors on those parameters,
our models remain resilient even in scenarios 
where hardware- or system-level defenses are not deployable.
Moreover, when combined with these existing defenses, 
the models lower space and computational complexities
by reducing the number of parameters to protect.

\textbf{Contributions.}
We \emph{first} present Hessian-aware training, 
a training algorithm designed to minimize the sharpness of a DNN's loss landscape,
thereby making the model less sensitive to parameter variations 
and more resilient to bitwise errors in its parameters.
While prior work proposed training algorithms
aimed at reducing the sharpness~\cite{%
    yao2021adahessian, foret2021sharpnessaware, yang2022hero},
we demonstrate that they are \emph{all} ineffective in 
enhancing a model's resilience to bitwise corruptions to its parameters.
Our training algorithm addresses this key problem,
and we further propose strategies to
make the training process computationally tractable 
when training ImageNet-scale models.

\emph{Second},
we conduct a comprehensive evaluation of 
our approach across multiple datasets and network architectures,
including those commonly used in prior studies.
We demonstrate that our training algorithm significantly enhances 
a model's resilience to bitwise errors to its parameters.
Models trained with our approach have 20--25\% of fewer parameters
where a single bitwise corruption causes a substantial accuracy drop 
compared to the baseline models.
Against multiple targeted bitwise corruptions,
our models require 2--3$\times$ more bit-flips 
to achieve the same malicious objectives.
Moreover, all models trained with our approach 
achieve accuracy the same as that of the baseline models.

\emph{Third}, 
we conduct an in-depth analysis of 
the increased resilience achieved by our approach.
Our analysis of visualized loss landscapes across layers
shows a great reduction in sharpness,
particularly in the layers close to the output.
Accordingly, we observe that the number of parameters 
where a single-bit corruption can cause a significant accuracy drop 
is mostly reduced in the fully-connected, classification layers.
Moreover, the numerical changes in parameter values 
required to cause the significant accuracy drop is increased.

\emph{Fourth},
we evaluate the compatibility of models 
trained with our training algorithm with existing defenses.
Our results demonstrate great synergy 
between our approach and prior work's defenses.
Fewer parameters are needed to be protected 
to achieve the same level of resilience observed 
when the baseline models are used.
Moreover, the defenses exhibit 
reduced runtime and storage overhead with our models.

\section{Background}
\label{sec:prelim}

\topic{IEEE-754 32-bit floating-point numbers}
are widely used to represent DNN parameters in memory. This format employs exponential notation and consists of three components: a sign bit that determines whether the number is positive or negative, the exponent (8 bits) that encodes the scale of the number using a biased representation, and the mantissa (23 bits) that encodes the precision. The significance of these components varies, with the exponent bits--particularly the most significant bits (MSBs)--having a disproportionately large impact on the represented value. For example, flipping the MSB of the exponent can cause drastic changes, such as turning a small value (e.g., 0.002) into an enormously large one (e.g., $6.8\times 10^{35}$). This characteristic makes protecting MSBs critical for ensuring numerical stability and resilience in DNNs. %
The majority of %
bits whose flipping causes a substantial accuracy drop 
are MSBs. In contrast, flipping bits in the mantissa typically results in minor perturbations %
with negligible impact on model performance.

\topic{Rowhammer attacks.}
Most bit-flip attacks on DNNs have demonstrated using Rowhammer,
a software-induced hardware fault-injection that exploits 
the physical structure of DRAM to induce bit-flips in memory~\cite{kim2014flipping}.
By repeatedly accessing (``hammering") specific memory rows, 
an attacker induces electrical disturbances in neighboring rows, 
causing bit-flips~\citep{di2023copy}.
These bit-flips can compromise data values in memory, 
such as DNN parameters, leading to severe dependability issues at runtime.
Rowhammer attacks have evolved from simple single-sided approaches 
to more sophisticated techniques like double-sided hammering, one-location hammering, 
and remote attacks via GPUs or network interfaces~\citep{zebram}, 
making them a versatile and persistent threat.
Rowhammer attackers can even target specific bits to flip%
~\citep{templating}. 
In DNNs, Rowhammer-induced bit-flips in critical bits, e.g., the MSBs of parameters, 
can lead to catastrophic changes to their behaviors, 
such as drastically reducing performance.
The widespread applicability of Rowhammer, 
requiring no physical access and exploiting shared resources in environments,
underscores its significance. %

\section{Our Hessian-aware Training}
\label{sec:method}

Now we design our training algorithm to 
enhance a model's resilience to parameter corruptions.
We focus on objectives that quantify a model's \emph{sensitivity}
to parameter value variations and use these objectives
as a loss function to minimize the sensitivity during training.
Suppose that a model $f$ uses a loss function $\mathcal{L}$.
The rate at which the loss changes in a specific variation direction $v$
within the parameter space can be expressed as the second-order derivative
${\partial^{2} \mathcal{L}}/{\partial v^{2}}$.
This value encodes how sensitive a model's performance will be
when its parameter values are changed along the direction of $v$.
During optimization, if the training algorithm minimizes this rate of change
across \emph{all} possible directions $v$,
the model will become resilient to parameter variations.

\subsection{The Hessian Trace as a Sensitivity Metric}
\label{subsec:metric}

\textbf{(\ul{Challenge 1})}
The next question becomes which metrics to use for 
capturing the sensitivity from the second-order derivatives.
A naive approach would compute the magnitude of the second-order derivatives
for a sufficiently large number of directions $v$ and average those values.
However, even with modern deep-learning frameworks like PyTorch,
which accelerate derivative computations,
computing numerous second-order derivatives of the loss 
at each mini-batch throughout training remains computationally intractable.

\textbf{(\ul{Challenge 2})}
Prior work has proposed various approaches to approximating the second-order derivatives%
~\citep{Jiang2020Fantastic, mulayoff2020unique, li2018visualizing, keskar2017on, neyshabur2017exploring}.
Our work utilizes the Hessian trace,
the sum of the eigenvalues of the Hessian matrix.
This also requires computing the Hessian matrix,
the second-order partial derivatives of 
a loss function, with respect to model parameters.
But there has been efficient methods we can use,
such as the Hutchinson's method~\citep{hutchinson} 
to approximate the Hessian trace efficiently over a number of random vectors $v$.
The key challenge is that,
while these studies mainly focus on using the approximated metrics
to measure the \emph{sharpness} (or flatness) of the loss landscape
and minimize it during training to improve DNN generalization,
their connection to model resilience to bitwise corruptions in parameters remains unknown.
A few studies have shown that minimizing the Hessian trace 
enhances model resilience to quantization---%
that induces optimal, bounded \emph{small} perturbations to parameters%
~\citep{hawq, hawqv2, hawqv3, yang2022hero}.
However, it is still unclear whether this resilience extends to
drastic, unbounded perturbations that bit-flips can induce.

\textbf{(\ul{Challenge 3})}
\citet{foret2021sharpnessaware} also noted that 
plugging-in the approximation as an objective into a standard numerical optimizer,
such as mini-batch stochastic gradient descent (SGD),
can result in instability during training.

\subsection{Minimizing the Hessian Trace in Training}
\label{subsec:practical-algorithms}

We present our training algorithm, designed to address the three aforementioned challenges
and reduce a model's sensitivity to its parameter value variations.
Our approach is to minimize Top-$p$ Hessian trace during training 
via an additional regularization term, %
outlined in Algorithm~\ref{algo:hessian_aware}.
The algorithm is an adaptation of the popular numerical optimizer, mini-batch SGD.
Notably, any gradient-based training methods
can be adapted to our Hessian-aware training.
The changes we made are highlighted in {\color{blue}blue}.

\begin{minipage}[t]{\linewidth}
\vspace{-2.em}
\begin{algorithm}[H]
\setstretch{0.90}
\caption{The Hessian-aware Training}
\label{algo:hessian_aware}
\textbf{Input:} 
    A model $f$, 
    Training data $D$,
    Training steps $T$,
    Learning rate $\eta$,
    Number of approximation steps %
    $p$,
    Regularization coefficient $\alpha$ \\
\textbf{Output:} A trained model $f_{\theta}$
\begin{algorithmic}[1]
\STATE Initialize $\theta_0$
{
\color{blue}
\STATE Initialize $\tau$ to 0
}
\FOR{$t=1,2, ..., T$}
    \STATE Draw a mini-batch $S_t$ from $D$
    \STATE Compute the loss $\mathcal{L}_{xe}(S_t; f_{\theta_t})$
    {
    \color{blue}
    \STATE $Tr_t, \lambda_t \gets 0, \phi$
    \FOR{$i=1,2, ..., p$}
        \STATE Draw a %
        vector $v_i$ %
        \STATE Compute the gradient $g_i$ of the loss $\mathcal{L}_{xe}$
        \STATE Compute the Hessian matrix $H_i$ along $v_i$
        \STATE Compute their eigenvalues $\lambda_i$ and trace $Tr_i$
        \STATE $Tr_t, \lambda_t \gets Tr_t + Tr_i, \lambda_t + \lambda_i$
    \ENDFOR
    \STATE $Tr_t, \lambda_t \gets (1/p) Tr_t, (1/p) \lambda_t$
    \IF{Median$(\lambda_t) > \tau$}           %
        \STATE $\mathcal{L}_{tot} \gets \mathcal{L}_{xe}(S_t; f_{\theta_t}) + \alpha * Tr_t$
    \ELSE
        \STATE $\mathcal{L}_{tot} \gets \mathcal{L}_{xe}(S_t; f_{\theta_t})$
        \STATE $\tau \gets $ Median($\lambda_t$)
    \ENDIF
    }
    \STATE Compute the gradient $g_t$ of $\mathcal{L}_{tot}$
    \STATE $\theta_{t+1} \gets \theta_t + \eta \cdot g_t$
\ENDFOR
\STATE \textbf{return} a trained model $f_{\theta}$
\end{algorithmic}
\end{algorithm}
\end{minipage}

In each mini-batch (line 3--22):

\textbf{(line 3--5, 20--21)}
We compute the loss $\mathcal{L}$ of a model $f_{\theta_t}$ 
and update the model parameters $\theta_{t}$ with its gradient $g_t$.
This step is the same as the original mini-batch SGD.

\textbf{(line 6--13)}
In this step, we compute the Hessian trace and eigenvalues with respect to the model parameters $\theta_{t}$.
Computing the full Hessian is computationally expensive; %
we approximate %
them using single step of the Hutchinson's method~\citep{hutchinson1989stochastic}, 
following the technique employed in prior work~\citep{yao2020pyhessian, yao2021adahessian}

Suppose the Hessian $H\!\in\!\mathbb{R}^{d\times d}$ 
and random vector $v\!\in\!\mathbb{R}^d$ 
satisfying $\mathbb{E}[vv^T]\!=\!\textbf{\emph{I}}$.
$v$ is drawn from Rademacher distribution 
which ensures half of the discrete probabilities are positive and the other half is negative ($P(v\!=\!\pm 1)\!=\!1/2$). 
$d$ denotes the total number of parameters.
In Hutchinson's method,
the Hessian trace %
over a set of random vectors is:
\begin{align*} \label{eq:trace_H}
    Tr(H) = \mathbb{E}[v^T Hv] = \frac{1}{p} \sum_{i=1}^{p} v^T_i Hv_i
\end{align*}
where $p$ is the number of random vectors used to approximate.
We can obtain $v^T Hv$ by computing the gradient of the loss function $\mathcal{L}$ twice as follows:
\begin{align*}
    v^T Hv = v^T \cdot \text{ } \frac{\partial }{\partial \theta} \Big ( \frac{\partial \mathcal{L}}{\partial \theta} \Big) \cdot v
\end{align*}

We follow the prior work~\citep{yao2020pyhessian} 
to compute set of Top-$p$ eigenvalues $\lambda{_p}$ as follows: 
\begin{align*}
    \lambda{_p} = \frac{v_i^T Hv_i}{\|v_i^T	\|} \quad \text{for} \, i = 1,2,\cdot \cdot \cdot \cdot, p
\end{align*}

\textbf{(line 14--19)}
In our experiments, we find that minimizing the Hessian trace computed on all eigenvalues 
(equal to the total number of parameters) renders the training process computationally intractable
as well as making the optimization process unstable. 
To address these issues,
we first take the $p$-largest eigenvalues to compute the loss.
There will be negligible impact 
since the eigenvalues consist of a few large values 
(representing the sharpest directions in the loss surface) 
and many smaller ones.
To identify an effective $p$ value,
we compare the effectiveness of %
choosing 10--50 eigenvalues in minimizing a model's sensitivity.

\begin{table}[ht]
\centering
\vspace{-1.em}
\caption{\textbf{Comparing our method using the Hessian trace from Top-$p$ eigenvalues.} 
Each row reports the average mean and standard deviations of the traces we compute over 1000 random samples, repeated five times across five different models.}
\vspace{0.2em}
\adjustbox{max width=\linewidth}{
    \begin{tabular}{@{}crr@{}}
    \toprule
    \textbf{$p$ value} & \multicolumn{1}{c}{\textbf{Test accuracy}} & \multicolumn{1}{r}{\textbf{Sensitivity}} \\ \midrule \midrule
    \textbf{Standard training} & 98.55 $\pm$ {\small0.53} & 128.58 $\pm$ {\small61.85}  \\ \midrule
    \textbf{Top-1 eigenvalue} & 98.37 $\pm$ {\small0.26} & 127.55 $\pm$ {\small34.51}  \\
    \textbf{Top-10 eigenvalues} & 98.16 $\pm$ {\small0.21} & 126.15 $\pm$ {\small63.59} \\
    \textbf{Top-25 eigenvalues} & 97.96 $\pm$ {\small0.22} & 116.10 $\pm$ {\small53.77} \\
    \textbf{Top-50 eigenvalues} & 98.92 $\pm$ {\small0.20} & 86.94 $\pm$ {\small38.93} \\ \bottomrule
    \end{tabular}
}
\label{tbl:topk-analysis}
\end{table}

Table~\ref{tbl:topk-analysis} summarizes our findings.
We train MNIST models and 
measure the sensitivity by computing the Hessian trace 
on a trained model.
We find that,
when we use Top-50 of the eigenvalues,
this results in the highest average accuracy of 98.92\%
and the lowest sensitivity (86.94\%).
We thus use the Top-50 eigenvalues for the rest of our paper.
To stabilize the training process,
we also track the trace values over the course of training 
and only regularize the model when the trace computed for a mini-batch 
is greater than the average trace values observed previously.
These two strategies we employ help stabilize our training
and allowing us to achieve reasonable performance and reduced model sensitivity.

\subsection{Evaluation Metrics}
\label{subsec:eval-metrics}

We introduce our evaluation metrics here to establish 
a clear framework for assessing our approach's effectiveness. 
In our evaluation of resilience against individual, single-bit corruptions~\cite{hong2019terminal},
we first define the \emph{distribution plot},
which counts the number of bits in a model's memory representation
that, when flipped, lead to the relative accuracy drop (RAD) specified on the x-axis.
RAD is defined as $(A_{c} - A_{p}) / A_{c}$,
where $A_{c}$ represents the classification accuracy of a model on a test set
and $A_{p}$ denotes the accuracy of the model under parameter corruptions.

\begin{figure}[ht]
\raggedright
\vspace{-0.8em}
\includegraphics[width=\linewidth, trim={0.5cm 0 0.3cm 0}, clip, keepaspectratio]{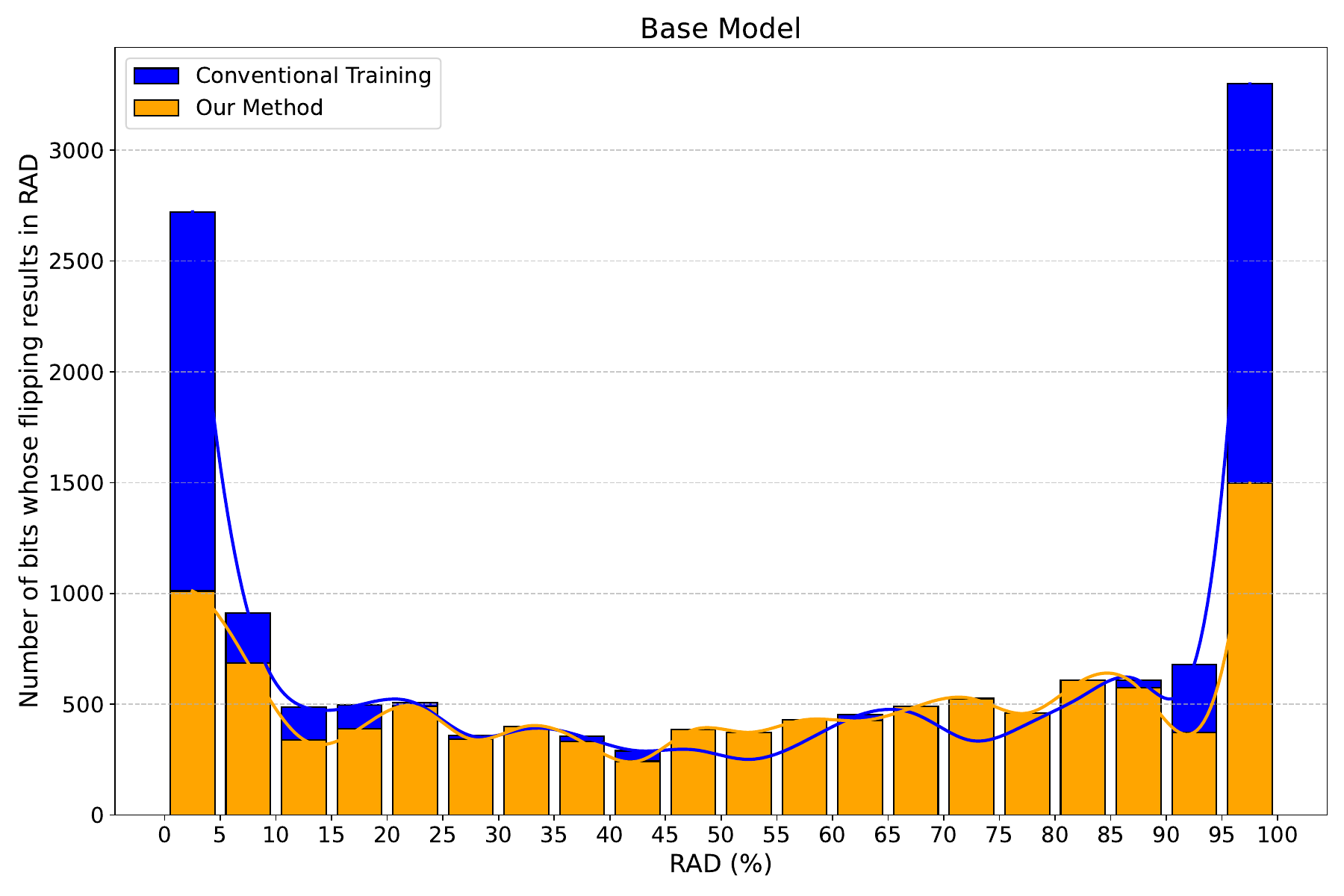}
\vspace{-1.5em}
\caption{\textbf{The distribution plots}
showing the number of bits in a DNN's parameters
whose flipping results in RAD on the x-axis.}
\label{fig:distribution-plot}
\end{figure}

Figure~\ref{fig:distribution-plot} shows example distribution plots
contrasting the two MNIST models, 
one trained with our training method and the other not.
We flip each bit individually and record RAD.
We use a 5\% granularity in RAD on the x-axis for our plots.
The plots show that our training method 
reduce the total number of bits whose flipping result in RADs
and also decrease the number of bit-flips leading to a 95–100\% RAD.
By using this plot, 
we gain a deeper understanding of 
the severity and impact of parameter perturbations 
before and after the application of our training algorithm.

We also define an \emph{erratic parameter}
as one where a single-bit error results in a RAD over 10\%.
We use this threshold 
because most prior work considers a 10\% RAD significant.

Moreover, we count the number of bit-flips required
to achieve a malicious objective,
such as complete accuracy depletion (RAD $>$ 90\%),
to assess our model's resilience against multi-bit corruption attacks%
~\cite{rakin2020tbt}.

\subsection{Comparison with Prior Approaches}
\label{subsec:our-effectiveness}

Next we empirically evaluate the effectiveness of our training approach 
compared to existing methods designed to train models with reduced sharpness.
We compare against three representative methods:
(1) $\ell_2$-regularization, which has been shown empirically
to reduce the sharpness of a model in literature~\citep{l2_flatness};
(2) AdaHessian~\citep{yao2021adahessian}, a second-order optimizer 
demonstrated to be effective in reducing the sharpness;
and (3) Sharpness-aware minimization (SAM)~\citep{foret2021sharpnessaware}, 
a training method specifically designed to reduce the sharpness.

\textbf{Methodology.}
We train MNIST and CIFAR-10 models and 
measure the accuracy and sensitivity. 
For each model, we compute the Hessian trace five times 
on 1000 randomly chosen training samples.
For each method, we run training five times and 
report the average. %
Because in MNIST, we find that 
SGD struggles with optimizing our second-order objective 
across the hyperparameters we use,
we use the RMSProp optimizer to benefit from a dynamic learning rate.
We choose the base learning rate and regularization coefficient 
$\alpha$ from \{1, 0.1, 0.01, 0.001, 0.0001\}, 
batch size from \{32, 64\}, 
and the number of Hutchinson's steps for trace approximation 
from \{1, 50, 100, 1000\}.
Through extensive hyper-parameter search, 
we find that using only a single step 
to compute the Hessian trace is %
the most effective.

\begin{table}[ht]
\centering
\vspace{-1.em}
\caption{\textbf{Comparison to existing training methods.} We compare the accuracy and the sensitivity from existing approaches to our method. The metrics are computed across five different models, and the sensitivity are computed over 1000 samples randomly chosen from the training data. We set $p$ to 50.}
\vspace{0.4em}
\adjustbox{max width=\linewidth}{
    \begin{tabular}{@{}l|cc|cc@{}}
\toprule
\multirow{2}{*}{\textbf{Training Method}} & \multicolumn{2}{c|}{\textbf{MNIST}} & \multicolumn{2}{c}{\textbf{CIFAR10}} \\ \cmidrule(l){2-5} 
 & \textbf{Acc.} & \textbf{Sensitivity} & \textbf{Acc.} & \textbf{Sensitivity} \\ \midrule \midrule
\textbf{Baseline} & 98.90 & 123.68 $\pm$ 63.79 & 92.43 & 3808.91 $\pm$ 803.19 \\
\textbf{L2-Regularization} & 97.30 & 128.23 $\pm$ 52.42 & 91.72 & 4117.33 $\pm$ 1032.42 \\
\textbf{AdaHessian} & 98.88 & 126.67 $\pm$ 70.82 & 92.68 & 3717.55 $\pm$ 931.80 \\
\textbf{SAM} & 97.15 & 134.08 $\pm$ 75.04 & 92.15 & 3676.89 $\pm$ 899.82 \\
\textbf{Ours} (Min-max) & 98.65 & 128.72 $\pm$ 68.50 & 92.34 & 3571.88 $\pm$ 924.67 \\
\textbf{Ours} (%
$\alpha$ to $10^{-4}$) & 98.78 & 126.67 $\pm$ 70.82 & 92.58 & 3543.33 $\pm$ 952.44 \\
\textbf{Ours} (%
$\alpha$ to 1) & 98.92 & 86.94 $\pm$ 38.93 & ---- & ---- \\ 
\textbf{Ours} (%
$\alpha$ to $10^{-2}$) & ---- & ---- & 92.71 & 2729.53 $\pm$ 762.94 \\
\bottomrule
\end{tabular}
}
\label{tbl:ours-vs-others}
\vspace{-0.4em}
\end{table}

\textbf{Results.}
Table~\ref{tbl:ours-vs-others} summarizes our results.
We show that compared to existing approaches,
our method is more effective in reducing a model's sensitivity.
We also employ two techniques
to smooth out the Hessian regularization loss $Tr_t$
that is fluctuating over training epochs:
(1) Min-max optimization: normalizing the loss 
based on the min and max values of the eigenvalues $\lambda_t$ as: 
$ Tr_{t_{norm}} = Tr_t - min (\lambda_t) / max(\lambda_t) - min(\lambda_t) $, 
where $Tr$ denotes the Hessian trace and $t$ denotes the current step;
and (2) the technique that only considers the loss
when its value is greater than 
the one $\tau$ observed in the previous epoch
(see line 14--19 in Algorithm~\ref{algo:hessian_aware}).
We additionally use this approach to determine and compare the impact of regularization coefficient $\alpha$,
and we show that setting $\alpha$ to one for MNIST and to $\alpha$ to $10^{-2}$ for CIFAR-10 achieves the lowest sensitivity.
For the rest of our experiments,
we use this training configurations.

\begin{table}[ht]
\centering
\vspace{-1.em}
\caption{\textbf{Contrasting our approach to existing second-order training methods.} BaseNet is trained on MNIST using AdaHessian, SAM, HERO~\cite{yang2022hero}, and our method. Column 4 reports the number of erratic parameters and column 6 their ratio to the total number of model parameters.}
\vspace{0.4em}
\adjustbox{max width=\linewidth}{
    \begin{tabular}{@{}l|c|c|c|c@{}}
    \toprule
    \textbf{Training Method} & \textbf{Acc.} & \textbf{\# Tot. Params} & \textbf{Erratic Params} & \textbf{Ratio} \\ \midrule \midrule
    \textbf{Baseline} & 98.73\% & \multirow{5}{*}{21,840} & 10,544 & 48.27\% \\ 
    \textbf{AdaHessian} & 98.88\% &  & 10,473 & 47.72\% \\ 
    \textbf{SAM} & 97.15\% &  & 10,621 & 48.63\% \\ 
    \textbf{HERO} & 98.27\% &  & 10,274 & 47.04\% \\ 
    \textbf{Ours} & 98.66\% &  & 8,482 & 38.83\% \\ \bottomrule
    \end{tabular}
}
\label{tbl:sam_adahessian_results}
\vspace{-0.4em}
\end{table}

Moreover, we evaluate the effectiveness of these training approaches in
enhacning a model's resilience to bitwise corruptions.
Table~\ref{tbl:sam_adahessian_results} compares the reduction in 
the number and ratio of erratic parameters for each method 
relative to the standard training baseline.
We also include a comparison with HERO~\cite{yang2022hero},
a method specifically designed to train models with guarantees
against bounded, small perturbations.
We find that all the existing approaches
reduce the number of erratic parameters by only a small margin (0--1.4\%), 
whereas ours achieves a 10\% reduction without any substantial accuracy drop.
Interestingly, HERO, %
designed to enhance resilience to bounded, small perturbations, 
offers only marginal resilience to bitwise corruptions.

\section{Empirical Evaluation}
\label{sec:eval}

We now comprehensively evaluate our training method.

\subsection{Experimental Setup}
\label{subsec:setup}

\textbf{Datasets.}
We use three image classification benchmarks:
MNIST~\citep{lecun2010mnist}, 
CIFAR-10~\citep{Krizhevsky09learningmultiple}, 
and ImageNet~\citep{ILSVRC15}.

\textbf{Models.}
We run our evaluation with four different DNNs.
For MNIST, we use two feed-forward DNNs: 
one with two convolutional layers and two fully-connected layers, 
and LeNet~\citep{lenet}.
For CIFAR-10 and ImageNet,
we consider DNN architectures popular in the community,
including ResNets~\citep{he2016deep} and a Transformer-based model,
DeiT-Tiny~\citep{touvron2021training}.

\textbf{Metrics.}
We employ the metrics introduced in \S\ref{subsec:eval-metrics}.
Against individual single-bit corruptions,
we compare the number of erratic parameters
and the total number of bits whose flipping results in
a specific RAD range, as illustrated in the distribution plot.
We also compare the number of bit-flips required to achieve a malicious objective
to assess the enhanced resilience to parameter corruptions.
Please refer to Appendix~\ref{appendix:experimental-setup} 
for details on our experimental setup.

\subsection{Quantifying Enhanced Model Resilience}
\label{subsec:results-on-bitflips}

We quantitatively analyze the resilience of models 
produced by our method to bitwise corruptions to its parameters.

\textbf{Methodology.}
We first evaluate our models against an adversary 
capable of inducing individual, single-bit corruptions in memory.
In most cases, practical fault-injection attacks like Rowhammer, 
are typically limited to flipping fewer than 10--20 bits in secure DRAM modules~\cite{jattke2022blacksmith},
this evaluation reflects a model's sensitivity
under the worst-case scenario of bitwise corruption (a single bit-flip).
We then extend our evaluation to adversaries 
capable of inducing multi-bit corruptions in memory~\cite{rakin2020tbt}.
Unlike the single-bit attackers, 
this scenario reflects the attacker continuously flipping bits in memory 
until they achieve a desirable accuracy drop, e.g., 90\% in RAD.

We individually test all 32 possible bit-flips 
in each model parameters for MNIST models (BaseNet and LeNet).
However, conducting the same analysis for CIFAR-10 and ImageNet models
is computationally infeasible, e.g., it takes 503 days 
to test all the bits in ResNet18 for CIFAR-10.
We thus employ the speed-up techniques proposed by~\citet{hong2019terminal}.
Because the bits most likely to cause substantial accuracy drops 
are the MSBs of the exponents, we focus on examining the exponent bits for CIFAR-10
and only the MSB of the exponents for ImageNet.
In ImageNet, for ResNet50, we test a randomly chosen 50\% of parameters
in all the convolutional layers and all the parameters in the fully-connected layers. We test all MSBs of the exponent of our transformer model.

\begin{table*}[ht]
\centering
\vspace{-0.8em}     %
\caption{\textbf{Enhanced resilience of models trained with our method.} We compare the resilience of models trained with our method (\emph{Ours}) to those trained without (\emph{Baseline}) against a single-bit error in their parameters.
\# Params are \# Bits are the total number of parameters and bits examined, and
\emph{Acc.} and \emph{Err.} refers to accuracy and erratic, respectively.
$\Delta$ is the reduction in erratic parameter ratios.}
\label{tab:model_vulnerability}
\vspace{0.2em}
\adjustbox{max width=0.9\linewidth}{
\begin{tabular}{@{}cccc|ccc|cccc@{}}
\toprule
\multirow{2}{*}{\textbf{Task}} & \multirow{2}{*}{\textbf{DNN}} & \multirow{2}{*}{\textbf{\# Params}} & \multirow{2}{*}{\textbf{\# Bits}} & \multicolumn{3}{c|}{\textbf{Baseline}} & \multicolumn{4}{c}{\textbf{Ours}} \\ \cmidrule(l){5-11} 
 &  &  &  & \textbf{Acc.} & \textbf{Err. Params} & \textbf{Ratio} & \textbf{Acc.} & \textbf{Err. Params} & \textbf{Ratio} & \textbf{$\Delta$} \\ \midrule \midrule
\multirow{2}{*}{\textbf{MNIST}} & \textbf{BaseNet} & 21,840 & 0.69M & 98.73 & 10,544 & 48.27\% & 98.66 & 8,482 & 38.83\% & -9.44\% \\
 & \textbf{LeNet} & 44,470 & 1.4M & 99.61 & 20,712 & 46.57\% & 98.91 & 15,383 & 34.59\% & -11.98\% \\ \midrule
\textbf{CIFAR-10} & \textbf{ResNet18} & 11M & 88M & 92.43 & 4.4M & 40.12\% & 93.68 & 3.7M & 33.6\% & -6.52\% \\ \midrule
\multirow{2}{*}{\textbf{ImageNet}} & \textbf{ResNet50} & 13.79M & 13.79M & 76.13 & 5.3M %
& 43.35\% & 75.09 & 4.5M %
& 36.59\% & -6.76\% \\ 
 & \textbf{DeiT-tiny} & 4.5M & 4.5M & 72.19 & 1.9M & 43.67\% & 71.93 & 1.6M & 36.84\% & -6.83\% \\ \bottomrule
\end{tabular}
}
\vspace{-0.4em}     %
\end{table*}

\textbf{Results.}
Table~\ref{tab:model_vulnerability} summarizes our results. 
We first note that our Hessian-aware training preserves model accuracy.
In all cases, the Acc. columns show that, 
there are negligible differences in Top-1 classification accuracy
between the baseline models and those trained with our method.
More importantly, our approach reduces 
the number of erratic parameters by 6.5--12.0\%:
In MNIST models, we observe a 10\% reduction,
while the reduction is 6.5--6.8\% in the CIFAR-10 and ImageNet models.
We attribute this difference in reductions
to the smaller number of erratic parameters in CIFAR-10 and ImageNet models,
which shows 3--8\% fewer erratic parameter ratios 
compared to MNIST models.
Surprisingly, for the ImageNet models,
even if we employ a training strategy
that fine-tunes only the last fully-connected layer,
(the most sensitive layer),
our method still enhances their resilience to individual, single-bit corruptions
by 6.8\%.

We also summarize our results on 
the multi-bit corruption adversary~\cite{rakin2019bit} 
in Table~\ref{tbl:targeted}.
This attack uses the progressive bit-search that
iteratively identifies the bit that maximizes performance degradation
while minimizing the number of bitwise errors needed.
To be consistent with the results from the original study, 
we evaluate our method on ImageNet models 
and use their attack configurations.
We report the number of bit-flips required to 
make the model to a random output generator (0.1\% accuracy).

\begin{table}[ht]
\centering
\vspace{-0.9em}
\caption{\textbf{Resilience of our models against multi-bit corruptions.}
We report the Top-1 accuracy of the models, along with 
the number of bit-flips required to reduce their accuracy to 0.1\%.} 
\vspace{0.2em}
\adjustbox{max width=0.84\linewidth}{
    \begin{tabular}{@{}ccccc@{}}
    \toprule
    \multirow{2}{*}{\textbf{Task}} & \multirow{2}{*}{\textbf{Model}} & \multirow{2}{*}{\textbf{Acc.}} 
        & \multicolumn{2}{c}{\textbf{\# Bit-flips Needed}} \\ \cmidrule{4-5}
     & & & \textbf{(Baseline)} & \textbf{(Ours)} \\ \midrule \midrule
    \multirow{2}{*}{\textbf{ImageNet}} & \textbf{ResNet18} & 69.57  & 13  & 31 \\
    & \textbf{ResNet50} & 75.33 & 11 & 29 \\ \bottomrule
    \end{tabular}
}

\label{tbl:targeted}
\vspace{-0.4em}
\end{table}

We show that models trained with our method require 
2--3$\times$ more bit-flips to achieve a 0.1\% accuracy.
The original study observes that 
DNNs with flatter loss landscapes requires more bit-flips
to degrade their performance significantly.
Our findings align with this observation; however,
our models exhibit an even flatter loss landscape than 
those examined in the study, which we believe contributes to 
their comparatively greater resilience against their attack.

\subsection{Characterization of the Enhanced Model Resilience}
\label{subsec:characterization}

We delve deeper into how various properties 
of a model influence its resilience to bitwise errors in parameters.

\begin{figure}[ht]
\raggedright
\includegraphics[width=\linewidth]{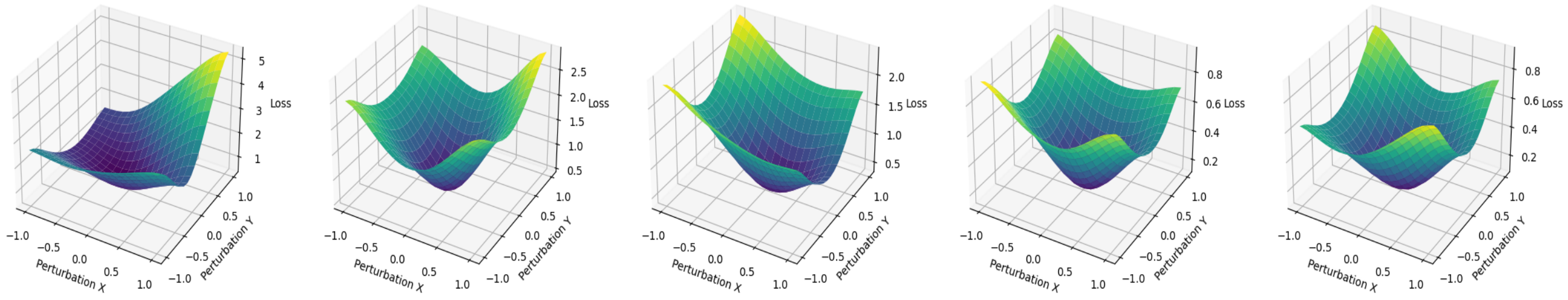}
\includegraphics[width=\linewidth]{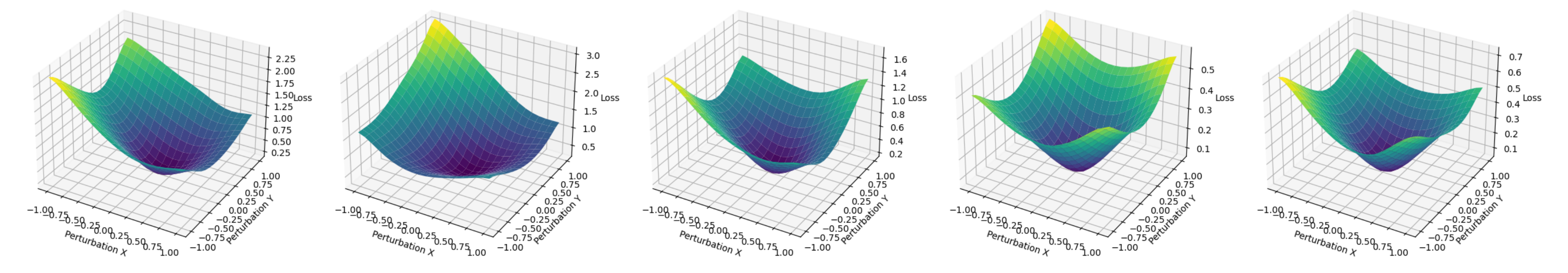}
\includegraphics[width=\linewidth]{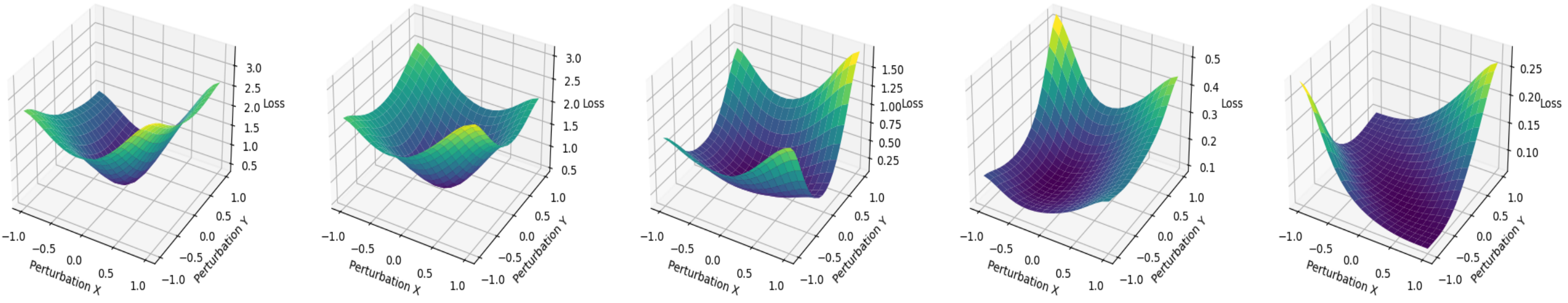}
\vspace{-0.8em}     %
\caption{\textbf{Visualizing LeNet's loss landscapes.}
From top to bottom, each row corresponds to standard training, HERO~\cite{yang2022hero}, and our method.
From left to right, we visualize the first two convolutional layers and the three fully-conected layers.}
\label{fig:visualization}
\vspace{-0.6em}     %
\end{figure}

\textbf{Visualizing the loss landscape.}
We first analyze whether the models trained with our method 
have a \emph{flatter} loss surface than the baselines.
We adopt the visualization technique 
proposed by~\citet{li2018visualizing}:
We choose two random vectors 
with the same dimension as that of a model's parameters
and incrementally increase the perturbations 
to each direction to the parameters
while measuring the loss value of the perturbed model.
Figure~\ref{fig:visualization} shows the loss landscape 
computed for each layer of the LeNet models trained on MNIST.
From left to right, we visualize the five layers from the input.

We show that our method effectively reduces the sharpness
across all layers, particularly for the layers close to the output.
In the last three columns of the figure,
which corresponds to the fully-connecte layers,
we observe that the loss curvatures become flatter
compared to other layers.
However, we also find that our approach is less effective
at reducing the sharpness of the layers close to the input
(the first two convolutional layers).
Model parameters closer to the output layers
are more prone to causing drastic changes in loss values
when subjected to variations.
Thus, our method focuses on minimizing the sharpness of these layers.

\begin{table*}[ht]
\centering
\vspace{-1em}     %
\caption{\textbf{Comparing the effectiveness of our approach in convolutional (Conv.) and fully-connected (FC) layers.} {Ours} refers to the models trained with our approach, while {Baseline} is to the models trained without. In Column 4, we show the \# of parameters in Conv or FC layers, with the parenthesis indicating their ratio in each model. All other numbers show erratic parameters and their ratios. The last two columns are the reduction in the two metrics. For ResNet50 conv layer, $^{\dagger}$ refers to 50\% sampled parameters.}
\label{tbl:layerwise-analysis}
\vspace{0.2em}
\begin{adjustbox}{width=0.92\textwidth}
\begin{tabular}{@{}cccr|cc|cc|cc@{}}
\toprule
\multirow{2}{*}{\textbf{Task}} & \multirow{2}{*}{\textbf{Model}} & \multirow{2}{*}{\textbf{Layers}} & \multicolumn{1}{c|}{\multirow{2}{*}{\textbf{\# Params}}} & \multicolumn{2}{c|}{\textbf{Baseline}} & \multicolumn{2}{c|}{\textbf{Ours}} & \multicolumn{2}{c}{\textbf{$\Delta$}} \\ \cmidrule(l){5-10} 
 &  &  & \multicolumn{1}{c|}{} & \textbf{Err. Params} & \textbf{Ratio} & \textbf{Err. Params} & \textbf{Ratio} & \textbf{Err. Params} & \textbf{Ratio} \\ \midrule \midrule
\multirow{4}{*}{\textbf{MNIST}} & \multirow{2}{*}{\textbf{BaseNet}} & \textbf{Conv.} & 5,280 (24.2\%) & 3,003 & 56.87\% & 2,695 & 51.04\% & -308 & -5.83\% \\
 &  & \textbf{FC} & 16,560 (75.8\%) & 7,544 & 45.55\% & 5,811 & 35.09\% & -1,733 & -10.46\% \\ \cmidrule{2-10}
 & \multirow{2}{*}{\textbf{LeNet}} & \textbf{Conv.} & 2,616 (5.9\%) & 1,719 & 65.71\% & 1,475 & 56.38\% & -244 & -9.33\% \\
 &  & \textbf{FC} & 41,854 (94.1\%) & 20,013 & 47.81\% & 14,903 & 35.61\% & -5,110 & -12.20\% \\ \midrule
\multirow{2}{*}{\textbf{CIFAR-10}} & \multirow{2}{*}{\textbf{ResNet18}} & \textbf{Conv.} & 11.2M (99.7\%) & 4.4M & 40.07\% & 3.7M & 33.57\% & -0.7M & -6.50\% \\
 &  & \textbf{FC} & 5,120 (0.03\%) & 2,297 & 44.86\% & 1,321 & 25.80\% & -976 & -19.06\% \\ \midrule
\multirow{2}{*}{\textbf{ImageNet}} & \multirow{2}{*}{\textbf{ResNet50}} & \textbf{Conv.} & $^{\dagger}$23.5M (53.5\%) & 4,516,162 & 38.23\% & 3,802,648 & 32.19\% & -713,514 & -6.04\% \\
 &  & \textbf{FC} & 2.04M (46.5\%) & 766,940 & 37.59\% & 656,493 & 32.18\% & -110,447 & -5.40\% \\ \bottomrule
\end{tabular}
\end{adjustbox}
\vspace{-0.5em}     %
\end{table*}

\textbf{Enhanced resilience in convolutional layers vs. fully-connected layers.}
Our previous analysis of the loss surfaces shows that 
the method tends to reduce the sensitivity 
(i.e., sharpness) of the later layers.
Since most feed-forward neural networks have 
convolutional layers followed by fully connected layers 
for classification, we analyze whether the resilience has 
indeed increased in the fully connected layers.
Table~\ref{tbl:layerwise-analysis} summarizes our findings.
Across all models, we observe that the reduction 
in the ratio of erratic parameters 
in fully connected layers is 2.4–13.4\% greater than 
that in convolutional layers.
Particularly, for the ResNet18 trained on CIFAR-10, 
our Hessian-aware training reduces 
the erratic parameter ratio by 19.1\%.
This result implies that network architectures 
with many fully connected layers, such as BaseNet or LeNet, 
can benefit more from our method.
However, architectures like ResNets, 
composed of 99\% of convolutional layers
followed by one or two fully connected layers, 
may experience a reduced benefit.

\begin{figure}[ht]
\raggedright
\vspace{-1.em}     %
\begin{minipage}{.49\linewidth}
    \centering
    \includegraphics[width=\linewidth, keepaspectratio]{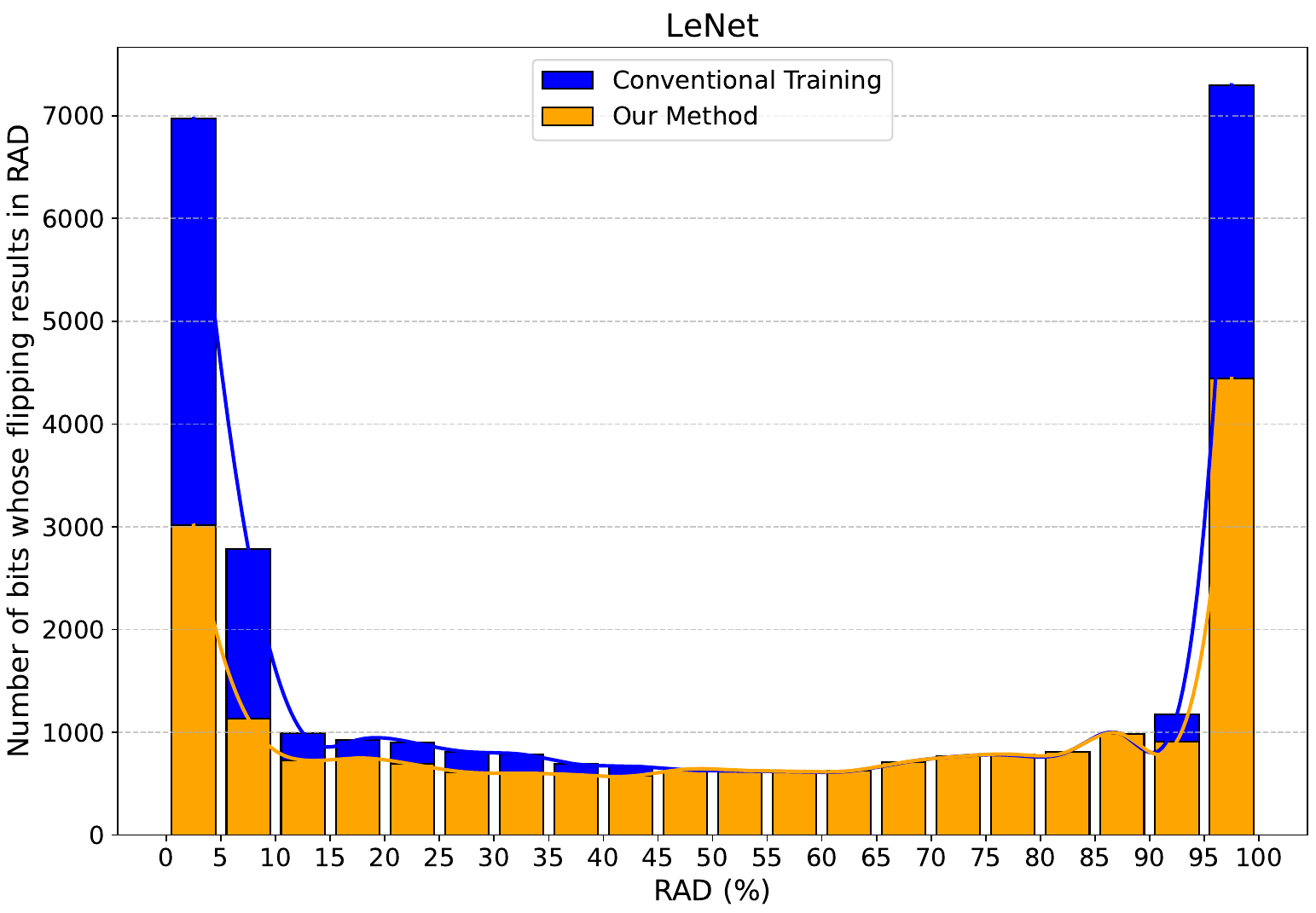}
\end{minipage}
\begin{minipage}{0.49\linewidth}
    \centering
    \includegraphics[width=\linewidth, keepaspectratio]{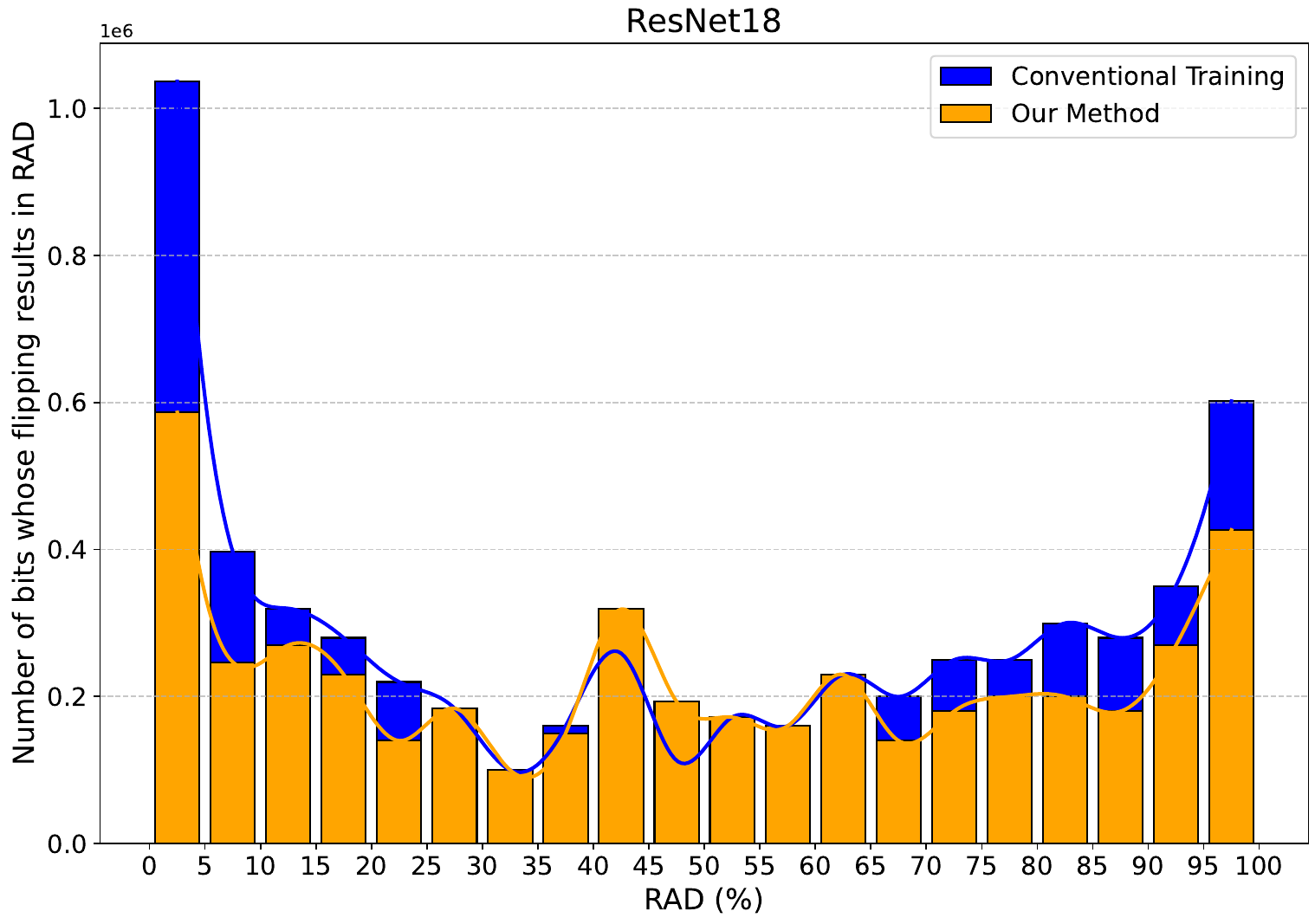}
    
\end{minipage}
\caption{%
\textbf{The distribution plots}
computed on LeNet in MNIST (left) and ResNet18 on CIFAR10 (right).}
\label{fig:distribution-plots}
\vspace{-0.5em}     %
\end{figure}

\textbf{Reduction in erratic bits.}
To gain a deeper insight into the enhanced resilience by our approach,
we contrast the distribution plots between two models:
one trained with our method or the other without.
Figure~\ref{fig:distribution-plots} shows the plots
from the LeNet (left) and ResNet18 (right) models.
Please see Appendix~\ref{appendix:imagenet_distribution}
for the plots from the ImageNet models.
In both models, our approach significantly reduces 
the number of erratic bits in two regions:
(1) bits whose flipping causes substantial performance loss (90--100\% RAD)
and (2) bits whose corruption result in small accuracy drops (0--10\% RAD).
This implies that our method reduces the likelihood 
of a model's performance degrading to a random output generator
by 25--50\%, caused by a single bit-flip.

\subsection{Synergy with Existing Defenses}
\label{subsec:existing-def}

In this section, we further demonstrate the synergy 
achieved by integrating system-level defenses proposed in the community 
with models trained with our method.

\topic{NeuroPot}~\citep{neuropots} injects honey (or decoy) neurons into a model
at locations likely to be targeted by an adversary,
without causing any significant accuracy drop.
Once these honey neurons are injected,
the defense needs system-level supports, such as an additional checksum modules
or a secure memory area, such as TEEs, to store the original parameters,
for detection and recovery.
Because our method reduces the number of erratic parameters, 
NeuroPot could benefit by requiring fewer honey neurons.
Table~\ref{tbl:neuropots} summarizes the benefits of
combining NeuroPot with models trained with our method.
We utilize LeNet for MNIST, ResNet18 for CIFAR-10, and ResNet50 for ImageNet. 

\begin{table}[ht]
\centering
\vspace{-1.em}
\caption{\textbf{Synergy with NeuroPot.}
We report the number of honey neurons (\# $h_n$) 
and the number of bit-flips (\# $bits$) required to 
cause an accuracy drop over 10\%, averaged over 5 runs.}
\vspace{0.2em}
\adjustbox{max width=\linewidth}{
    \begin{tabular}{@{}c|c|c|c|c|c|c|c|c@{}}
    \toprule
    \multirow{2}{*}{\textbf{Model}} & \multicolumn{3}{c|}{\textbf{Baseline}} & \multicolumn{5}{c}{\textbf{Ours}} \\ \cmidrule{2-9}
     & \textbf{Acc.} & \# $h_n$ & \# bits & \textbf{Acc.} & \# $h_n$ & \# bits & \# $h_n$ & \# bits \\ \midrule \midrule
    \textbf{LeNet}  & 99.32  & 25  & 11 $\pm$ 2 & 99.47 & 25 & 21 $\pm$ 4 & 10 & 12 $\pm$ 2 \\
    \textbf{ResNet18}  & 92.17  & 50 & 15 $\pm$ 3 & 92.13 & 50 & 34 $\pm$  3 & 20 & 15 $\pm$4 \\ 
    \textbf{ResNet50} & 76.09 & 150  & 17 $\pm$ 3 & 76.01 & 150 & 31 $\pm$ 5 & 30 & 16 $\pm$ 4\\ \bottomrule
    \end{tabular}
}
\label{tbl:neuropots}
\vspace{-0.4em}
\end{table}

NeuroPot, when combined with our method, 
enhances the resilience or reduces space complexity.
If the number of honey neurons (\# $h_n$) is fixed, 
models trained with our method require twice as many bit-flips (\# $bits$) 
to cause a 10\% accuracy drop compared to the baselines.
If an adversary is allowed to cause a 10\% accuracy drop
with the same \# $bits$, our models need 60--80\% fewer honey neurons,
thereby improving inference time and storage efficiency.
For instance, reducing to 30 $h_n$ in the ResNet50 ImageNet model 
decreases inference time by 55.9\% (0.37s vs. 0.84s) 
and storage overhead by 65\% (34KB vs. 99KB).

\topic{RADAR~\citep{radar}} is a checksum-based defense,
which stores \emph{golden signature} for a group of weights
and compares this signature at runtime 
with the current model signature.
We adapt this scheme to store the golden signature of erratic parameters,
further enhancing resilience at runtime.
In our evaluation, we use two models:
ResNet20 for CIFAR-10 and ResNet18 for ImageNet.
We use the group sizes specified in the original study%
---8 for ResNet20 and 512 for ResNet18.
Table~\ref{tbl:radar} shows our results.

\begin{table}[!ht]
\centering
\vspace{-0.8em}
\caption{\textbf{Synergy with RADAR.}
We report the accuracy after 15 random bit-flips (3rd and 5th columns) 
and the accuracy after the recovery using RADAR (4th and 6th columns).}
\vspace{0.4em}
\adjustbox{max width=0.9\linewidth}{
    \begin{tabular}{@{}cc|cc|cc@{}}
    \toprule
    \multirow{2.5}{*}{\textbf{Model}} &  \multirow{2.5}{*}{\makecell{\textbf{Initial} \\ 
    \textbf{Acc.}}} & \multicolumn{2}{c}{\textbf{Baseline}} & \multicolumn{2}{|c}{\textbf{Ours}} \\ \cmidrule{3-6} 
     &  & 
     \textbf{Acc.} & \textbf{Recovery} & \textbf{Acc.} & \textbf{Recovery}  \\ \midrule \midrule
     \textbf{ResNet20} & 90.13  & 18.01  & 81.13 & 27.93 & 88.23 \\ 
     \textbf{ResNet18} & 69.34 & 0.19 & 60.18 & 15.29 & 64.88 \\ \bottomrule
    \end{tabular}
}
\label{tbl:radar}
\vspace{-0.8em}
\end{table}

We demonstrate the synergy of our method when combined with RADAR.
With RADAR, models trained with our method achieve better accuracy recovery%
--88.23\% for ResNet20 and 64.88\% for ResNet18--compared to baseline models,
which recover 81.13\% and 60.18\%, respectively.

\begin{table}[!ht]
\centering
\vspace{-0.8em}
\caption{\textbf{Impact of RADAR configurations on runtime and space complexity.} 
G denotes the group size and Acc. Rec. refers to the accuracy recovery achieved using RADAR scheme.
Inference time is measured in milliseconds (ms), and space complexity is measured in kilobytes (kB).}
\vspace{0.4em}
\adjustbox{max width=\linewidth}{
    \begin{tabular}{@{}c|cccc|cccc@{}}
    \toprule
    \multirow{2.5}{*}{\textbf{Model}} &  \multicolumn{4}{c}{\textbf{Baseline}} & \multicolumn{4}{|c}{\textbf{Ours}} \\ \cmidrule{2-9}
     & \textbf{G} & \makecell{\textbf{Acc.} \\ \textbf{Rec.}} & \textbf{Time} & \textbf{Space} & \textbf{G} & \makecell{\textbf{Acc.} \\ \textbf{Rec.}} & \textbf{Time} & \textbf{Space}  \\ \midrule \midrule
     \textbf{ResNet20} & 8 & 81.13 & 0.06ms & 8.2kB & 64 & 80.93 & 0.02ms & 3.2kB \\ 
     \textbf{ResNet18} & 512 & 60.18 & 3.32ms & 5.6kB & 1024 & 59.47  & 1.86ms & 2.95kB \\ \bottomrule
    \end{tabular}
}
\label{tbl:radar_imp}
\vspace{-0.4em}
\end{table}

Our method further enhances the inference overhead and space complexity.
Table~\ref{tbl:radar_imp} shows our results.
For example, on ResNet20 trained on CIFAR-10, 
increasing the group size by 4$\times$ to 64
results in a reduced accuracy recovery of 80.93\%, 
which is still comparable to the baseline model's runtime recovery (81.13\%).
But, the choice offers significant benefits:
a 69\% reduction in both runtime and space complexity
compared to the baseline CIFAR-10 model.

\section{Discussion}
\label{sec:discussion}

\textbf{Increase in computational demands.} We evaluate the overhead of our training method in terms of actual training wall-time measured in PyTorch on a NVIDIA Tesla V100 GPU. In Appendix~\ref{appendix:overhead} we present our results. The Hessian-aware training incurs overhead that scales with the size of the model; a 4--6$\times$ times increase in computations for MNIST models, and a 10$\times$ times increase in overhead for CIFAR-10 models. Existing works utilizing second-order properties during training take a completely different approach compute the Hessian and its eigenvalues: they employ weight perturbations~\citep{foret2021sharpnessaware} or only the trace approximation~\citep{yao2021adahessian} to minimize sharpness of the loss landscape. The increase in computation in our approach is primarily attributed to the large Hessian and eigenvalues we need to compute with respect to model parameters, which is not optimized for popular deep-learning frameworks. To reduce the computational overhead during training, we employ a layer sampling technique. As prior work identifies the last layers to be most susceptible to bitwise errors~\citep{hong2019terminal}, we believe only computing Hessian trace on last few layers can aid resilient model training. Our results for large-scale models, such as ResNet50 in ImageNet, show that this technique significantly reduces the computational overhead from 10$\times$ times to 1.18$\times$ times, being equally effective in enhancing model resilience. %

\textbf{Hardware-level defenses.} Many hardware-level defenses are designed to mitigate RowHammer~\citep{kim2014flipping}, a software-induced attack that causes a targeted DRAM row to leak capacitance by repeatedly accessing its neighboring rows. \citet{kim2014flipping} have proposed a
defense that proactively refreshes rows that are frequently accessed, as they are at higher risk of being targeted by the attack. Panopticon~\citep{bennett2021panopticon} leverages a similar idea: it employs hardware counters for each data row in DRAM and refreshes the rows when the counter reaches a predefined threshold. Instead of refreshing the rows at high risks, \citet{saileshwar2022randomized} propose swapping
them with safe memory regions. \citet{di2023copy} use the error correction codes as a mechanism for triggering such swapping. DRAM-Locker~\citep{zhou2023dram} leverages a lock-table in SRAM. If the addresses of the high-risk rows are stored in the lock-table, any access this addresses without the unlock command will be denied. %
These defenses mainly protect data rows at high risk of being targeted. 
Our work reduces the number of data rows in a model whose perturbations lead to
significant accuracy loss, and therefore, potentially decreasing their runtime and space overheads.
However, we note that existing hardware-level defenses require infrastructure changes
or additional hardware components. %

\section{Conclusion}
\label{sec:conclusion}

Our work presents a training algorithm 
designed to reduce a model's sensitivity to parameter variations, 
thereby enhancing its resilience to
bitwise corruptions in model parameters.
We focus on the model's second-order property, 
the Hessian trace, and develop an objective function 
to directly minimize it during training.
We extensively compare our approach 
with existing methods for improving model resilience 
and demonstrate its effectiveness.
We evaluate our approach by testing 
a model's performance under both single-bit and multi-bit 
parameter corruptions in memory.
Our method reduces the number of erratic parameters by 6--12\%, 
and decreases those causing a 90--100\% RAD drop by 20--50\%.
We also increase the number of bit-flips
required by a multi-bit adversary
to induce substantial accuracy drops.
Moreover, we demonstrate the synergy 
when combined with system-level defenses
to protecting models against parameter-corruption attacks.
We hope our work will inspire future work on
the safe deployment of deep neural networks in emerging computing platforms.

\section*{Acknowledgement}

This work is partially supported by the Samsung Global Research Outreach (GRO) program.
The findings and conclusions in this work are those of the author(s) 
and do not necessarily represent the views of the funding agency.

\section*{Impact Statement.}

This paper presents a hessian-based DNN training approach to enhancing the inherent resilience of a DNN model to unbounded bitwise errors frequently observed in emerging hardware devices. By exploring Hessian-based training and evaluating their efficacy across various datasets and architectures, this work significantly advances understanding and mitigation of hardware-induced vulnerabilities in DNNs. The proposed methodology not only improves the resilience of the models but also bridges critical gaps in deploying machine learning systems in error-prone devices. This research has potential implications for industries relying on error-prone hardware where resilience to hardware perturbations is paramount. By ensuring reliable performance under challenging conditions, our work contributes to the broader adoption of DNNs across diverse fields, fostering innovation and reliability in next-generation computing systems.

{
    \bibliographystyle{unsrtnat}
    \bibliography{bib/references}

\begin{thebibliography}{48}
\providecommand{\natexlab}[1]{#1}
\providecommand{\url}[1]{\texttt{#1}}
\expandafter\ifx\csname urlstyle\endcsname\relax
  \providecommand{\doi}[1]{doi: #1}\else
  \providecommand{\doi}{doi: \begingroup \urlstyle{rm}\Url}\fi

\bibitem[Hong et~al.(2019)Hong, Frigo, Kaya, Giuffrida, and
  Dumitraș]{hong2019terminal}
Sanghyun Hong, Pietro Frigo, Yi{\u{g}}itcan Kaya, Cristiano Giuffrida, and
  Tudor Dumitraș.
\newblock Terminal brain damage: Exposing the graceless degradation in deep
  neural networks under hardware fault attacks.
\newblock In \emph{28th USENIX Security Symposium (USENIX Security 19)}, pages
  497--514, 2019.

\bibitem[Rakin et~al.(2019)Rakin, He, and Fan]{rakin2019bit}
Adnan~Siraj Rakin, Zhezhi He, and Deliang Fan.
\newblock Bit-flip attack: Crushing neural network with progressive bit search.
\newblock In \emph{Proceedings of the IEEE/CVF International Conference on
  Computer Vision}, pages 1211--1220, 2019.

\bibitem[Yao et~al.(2020{\natexlab{a}})Yao, Rakin, and Fan]{deephammer}
Fan Yao, Adnan~Siraj Rakin, and Deliang Fan.
\newblock {DeepHammer}: Depleting the intelligence of deep neural networks
  through targeted chain of bit flips.
\newblock In \emph{29th USENIX Security Symposium (USENIX Security 20)}, pages
  1463--1480. USENIX Association, August 2020{\natexlab{a}}.
\newblock ISBN 978-1-939133-17-5.
\newblock URL
  \url{https://www.usenix.org/conference/usenixsecurity20/presentation/yao}.

\bibitem[Bai et~al.(2023)Bai, Wu, Li, and Xia]{bai2023versatile}
Jiawang Bai, Baoyuan Wu, Zhifeng Li, and Shu-Tao Xia.
\newblock Versatile weight attack via flipping limited bits.
\newblock \emph{IEEE Transactions on Pattern Analysis and Machine
  Intelligence}, 2023.

\bibitem[Cai et~al.(2021)Cai, Chowdhuryy, Zhang, and Yao]{cai2021seeds}
Kunbei Cai, Md~Hafizul~Islam Chowdhuryy, Zhenkai Zhang, and Fan Yao.
\newblock Seeds of seed: Nmt-stroke: Diverting neural machine translation
  through hardware-based faults.
\newblock In \emph{2021 International Symposium on Secure and Private Execution
  Environment Design (SEED)}, pages 76--82. IEEE, 2021.

\bibitem[Rakin et~al.(2021{\natexlab{a}})Rakin, He, Li, Yao, Chakrabarti, and
  Fan]{rakin2021t}
Adnan~Siraj Rakin, Zhezhi He, Jingtao Li, Fan Yao, Chaitali Chakrabarti, and
  Deliang Fan.
\newblock T-bfa: Targeted bit-flip adversarial weight attack.
\newblock \emph{IEEE Transactions on Pattern Analysis and Machine
  Intelligence}, 44\penalty0 (11):\penalty0 7928--7939, 2021{\natexlab{a}}.

\bibitem[Chen et~al.(2021)Chen, Fu, Zhao, and Koushanfar]{Chen_2021_ICCV}
Huili Chen, Cheng Fu, Jishen Zhao, and Farinaz Koushanfar.
\newblock Proflip: Targeted trojan attack with progressive bit flips.
\newblock In \emph{Proceedings of the IEEE/CVF International Conference on
  Computer Vision (ICCV)}, pages 7718--7727, October 2021.

\bibitem[Cai et~al.(2024)Cai, Chowdhuryy, Zhang, and Yao]{10646710}
Kunbei Cai, Md~Hafizul~Islam Chowdhuryy, Zhenkai Zhang, and Fan Yao.
\newblock { DeepVenom: Persistent DNN Backdoors Exploiting Transient Weight
  Perturbations in Memories }.
\newblock In \emph{2024 IEEE Symposium on Security and Privacy (SP)}, pages
  2067--2085, Los Alamitos, CA, USA, May 2024. IEEE Computer Society.
\newblock \doi{10.1109/SP54263.2024.00223}.
\newblock URL
  \url{https://doi.ieeecomputersociety.org/10.1109/SP54263.2024.00223}.

\bibitem[Rakin et~al.(2020)Rakin, He, and Fan]{rakin2020tbt}
Adnan~Siraj Rakin, Zhezhi He, and Deliang Fan.
\newblock Tbt: Targeted neural network attack with bit trojan.
\newblock In \emph{Proceedings of the IEEE/CVF Conference on Computer Vision
  and Pattern Recognition}, pages 13198--13207, 2020.

\bibitem[Rakin et~al.(2022)Rakin, Chowdhuryy, Yao, and Fan]{deepsteal}
Adnan~Siraj Rakin, Md~Hafizul~Islam Chowdhuryy, Fan Yao, and Deliang Fan.
\newblock Deepsteal: Advanced model extractions leveraging efficient weight
  stealing in memories.
\newblock In \emph{2022 IEEE Symposium on Security and Privacy (SP)}, pages
  1157--1174, 2022.
\newblock \doi{10.1109/SP46214.2022.9833743}.

\bibitem[Bennett et~al.(2021)Bennett, Saroiu, Wolman, and
  Cojocar]{bennett2021panopticon}
Tanj Bennett, Stefan Saroiu, Alec Wolman, and Lucian Cojocar.
\newblock Panopticon: A complete in-dram rowhammer mitigation.
\newblock In \emph{Workshop on DRAM Security (DRAMSec)}, volume~22, page 110,
  2021.

\bibitem[Rakin et~al.(2021{\natexlab{b}})Rakin, Yang, Li, Yao, Chakrabarti,
  Cao, Seo, and Fan]{rakin2021ra}
Adnan~Siraj Rakin, Li~Yang, Jingtao Li, Fan Yao, Chaitali Chakrabarti, Yu~Cao,
  Jae-sun Seo, and Deliang Fan.
\newblock Ra-bnn: Constructing robust \& accurate binary neural network to
  simultaneously defend adversarial bit-flip attack and improve accuracy.
\newblock \emph{arXiv preprint arXiv:2103.13813}, 2021{\natexlab{b}}.

\bibitem[Li et~al.(2021)Li, Rakin, He, Fan, and Chakrabarti]{radar}
Jingtao Li, Adnan~Siraj Rakin, Zhezhi He, Deliang Fan, and Chaitali
  Chakrabarti.
\newblock Radar: Run-time adversarial weight attack detection and accuracy
  recovery.
\newblock In \emph{2021 Design, Automation \& Test in Europe Conference \&
  Exhibition (DATE)}, pages 790--795, 2021.
\newblock \doi{10.23919/DATE51398.2021.9474113}.

\bibitem[Di~Dio et~al.(2023)Di~Dio, Koning, Bos, and Giuffrida]{di2023copy}
Andrea Di~Dio, Koen Koning, Herbert Bos, and Cristiano Giuffrida.
\newblock Copy-on-flip: Hardening ecc memory against rowhammer attacks.
\newblock In \emph{NDSS}, 2023.

\bibitem[Zhou et~al.(2023{\natexlab{a}})Zhou, Ahmed, Rakin, and
  Angizi]{zhou2023dnn}
Ranyang Zhou, Sabbir Ahmed, Adnan~Siraj Rakin, and Shaahin Angizi.
\newblock Dnn-defender: An in-dram deep neural network defense mechanism for
  adversarial weight attack.
\newblock \emph{arXiv preprint arXiv:2305.08034}, 2023{\natexlab{a}}.

\bibitem[Zhou et~al.(2023{\natexlab{b}})Zhou, Ahmed, Roohi, Rakin, and
  Angizi]{zhou2023dram}
Ranyang Zhou, Sabbir Ahmed, Arman Roohi, Adnan~Siraj Rakin, and Shaahin Angizi.
\newblock Dram-locker: A general-purpose dram protection mechanism against
  adversarial dnn weight attacks.
\newblock \emph{arXiv preprint arXiv:2312.09027}, 2023{\natexlab{b}}.

\bibitem[Liu et~al.(2023)Liu, Yin, Wen, Yang, and Sha]{neuropots}
Qi~Liu, Jieming Yin, Wujie Wen, Chengmo Yang, and Shi Sha.
\newblock {NeuroPots}: Realtime proactive defense against {Bit-Flip} attacks in
  neural networks.
\newblock In \emph{32nd USENIX Security Symposium (USENIX Security 23)}, pages
  6347--6364, Anaheim, CA, August 2023. USENIX Association.
\newblock ISBN 978-1-939133-37-3.
\newblock URL
  \url{https://www.usenix.org/conference/usenixsecurity23/presentation/liu-qi}.

\bibitem[Wang et~al.(2023)Wang, Zhang, Wang, Qiu, Zhang, Li, Li, Wei, and
  Zhang]{ageis}
Jialai Wang, Ziyuan Zhang, Meiqi Wang, Han Qiu, Tianwei Zhang, Qi~Li, Zongpeng
  Li, Tao Wei, and Chao Zhang.
\newblock Aegis: Mitigating targeted bit-flip attacks against deep neural
  networks.
\newblock In \emph{32nd USENIX Security Symposium (USENIX Security 23)}, pages
  2329--2346, Anaheim, CA, August 2023. USENIX Association.
\newblock ISBN 978-1-939133-37-3.
\newblock URL
  \url{https://www.usenix.org/conference/usenixsecurity23/presentation/wang-jialai}.

\bibitem[Yao et~al.(2021{\natexlab{a}})Yao, Gholami, Shen, Mustafa, Keutzer,
  and Mahoney]{yao2021adahessian}
Zhewei Yao, Amir Gholami, Sheng Shen, Mustafa Mustafa, Kurt Keutzer, and
  Michael Mahoney.
\newblock Adahessian: An adaptive second order optimizer for machine learning.
\newblock In \emph{proceedings of the AAAI conference on artificial
  intelligence}, volume~35, pages 10665--10673, 2021{\natexlab{a}}.

\bibitem[Foret et~al.(2021)Foret, Kleiner, Mobahi, and
  Neyshabur]{foret2021sharpnessaware}
Pierre Foret, Ariel Kleiner, Hossein Mobahi, and Behnam Neyshabur.
\newblock Sharpness-aware minimization for efficiently improving
  generalization.
\newblock In \emph{International Conference on Learning Representations}, 2021.
\newblock URL \url{https://openreview.net/forum?id=6Tm1mposlrM}.

\bibitem[Yang et~al.(2022)Yang, Yang, Gong, and Chen]{yang2022hero}
Huanrui Yang, Xiaoxuan Yang, Neil~Zhenqiang Gong, and Yiran Chen.
\newblock Hero: Hessian-enhanced robust optimization for unifying and improving
  generalization and quantization performance.
\newblock In \emph{Proceedings of the 59th ACM/IEEE Design Automation
  Conference}, pages 25--30, 2022.

\bibitem[Kim et~al.(2014)Kim, Daly, Kim, Fallin, Lee, Lee, Wilkerson, Lai, and
  Mutlu]{kim2014flipping}
Yoongu Kim, Ross Daly, Jeremie Kim, Chris Fallin, Ji~Hye Lee, Donghyuk Lee,
  Chris Wilkerson, Konrad Lai, and Onur Mutlu.
\newblock Flipping bits in memory without accessing them: An experimental study
  of dram disturbance errors.
\newblock \emph{ACM SIGARCH Computer Architecture News}, 42\penalty0
  (3):\penalty0 361--372, 2014.

\bibitem[Konoth et~al.(2018)Konoth, Oliverio, Tatar, Andriesse, Bos, Giuffrida,
  and Razavi]{zebram}
Radhesh~Krishnan Konoth, Marco Oliverio, Andrei Tatar, Dennis Andriesse,
  Herbert Bos, Cristiano Giuffrida, and Kaveh Razavi.
\newblock {ZebRAM}: Comprehensive and compatible software protection against
  rowhammer attacks.
\newblock In \emph{13th USENIX Symposium on Operating Systems Design and
  Implementation (OSDI 18)}, pages 697--710, Carlsbad, CA, October 2018. USENIX
  Association.
\newblock ISBN 978-1-939133-08-3.
\newblock URL
  \url{https://www.usenix.org/conference/osdi18/presentation/konoth}.

\bibitem[Razavi et~al.(2016)Razavi, Gras, Bosman, Preneel, Giuffrida, and
  Bos]{templating}
Kaveh Razavi, Ben Gras, Erik Bosman, Bart Preneel, Cristiano Giuffrida, and
  Herbert Bos.
\newblock Flip feng shui: Hammering a needle in the software stack.
\newblock In \emph{25th USENIX Security Symposium (USENIX Security 16)}, pages
  1--18, 2016.

\bibitem[Jiang et~al.(2020)Jiang, Neyshabur, Mobahi, Krishnan, and
  Bengio]{Jiang2020Fantastic}
Yiding Jiang, Behnam Neyshabur, Hossein Mobahi, Dilip Krishnan, and Samy
  Bengio.
\newblock Fantastic generalization measures and where to find them.
\newblock In \emph{International Conference on Learning Representations}, 2020.
\newblock URL \url{https://openreview.net/forum?id=SJgIPJBFvH}.

\bibitem[Mulayoff and Michaeli(2020)]{mulayoff2020unique}
Rotem Mulayoff and Tomer Michaeli.
\newblock Unique properties of flat minima in deep networks.
\newblock In \emph{International conference on machine learning}, pages
  7108--7118. PMLR, 2020.

\bibitem[Li et~al.(2018)Li, Xu, Taylor, Studer, and
  Goldstein]{li2018visualizing}
Hao Li, Zheng Xu, Gavin Taylor, Christoph Studer, and Tom Goldstein.
\newblock Visualizing the loss landscape of neural nets.
\newblock \emph{Advances in neural information processing systems}, 31, 2018.

\bibitem[Keskar et~al.(2017)Keskar, Mudigere, Nocedal, Smelyanskiy, and
  Tang]{keskar2017on}
Nitish~Shirish Keskar, Dheevatsa Mudigere, Jorge Nocedal, Mikhail Smelyanskiy,
  and Ping Tak~Peter Tang.
\newblock On large-batch training for deep learning: Generalization gap and
  sharp minima.
\newblock In \emph{International Conference on Learning Representations}, 2017.
\newblock URL \url{https://openreview.net/forum?id=H1oyRlYgg}.

\bibitem[Neyshabur et~al.(2017)Neyshabur, Bhojanapalli, McAllester, and
  Srebro]{neyshabur2017exploring}
Behnam Neyshabur, Srinadh Bhojanapalli, David McAllester, and Nati Srebro.
\newblock Exploring generalization in deep learning.
\newblock \emph{Advances in neural information processing systems}, 30, 2017.

\bibitem[Bekas et~al.(2007)Bekas, Kokiopoulou, and Saad]{hutchinson}
C.~Bekas, E.~Kokiopoulou, and Y.~Saad.
\newblock An estimator for the diagonal of a matrix.
\newblock \emph{Appl. Numer. Math.}, 57\penalty0 (11–12):\penalty0
  1214–1229, nov 2007.
\newblock ISSN 0168-9274.
\newblock \doi{10.1016/j.apnum.2007.01.003}.
\newblock URL \url{https://doi.org/10.1016/j.apnum.2007.01.003}.

\bibitem[Dong et~al.(2019)Dong, Yao, Gholami, Mahoney, and Keutzer]{hawq}
Z.~Dong, Z.~Yao, A.~Gholami, M.~Mahoney, and K.~Keutzer.
\newblock Hawq: Hessian aware quantization of neural networks with
  mixed-precision.
\newblock In \emph{2019 IEEE/CVF International Conference on Computer Vision
  (ICCV)}, pages 293--302, Los Alamitos, CA, USA, nov 2019. IEEE Computer
  Society.
\newblock \doi{10.1109/ICCV.2019.00038}.
\newblock URL
  \url{https://doi.ieeecomputersociety.org/10.1109/ICCV.2019.00038}.

\bibitem[Dong et~al.(2020)Dong, Yao, Arfeen, Gholami, Mahoney, and
  Keutzer]{hawqv2}
Zhen Dong, Zhewei Yao, Daiyaan Arfeen, Amir Gholami, Michael~W Mahoney, and
  Kurt Keutzer.
\newblock Hawq-v2: Hessian aware trace-weighted quantization of neural
  networks.
\newblock In H.~Larochelle, M.~Ranzato, R.~Hadsell, M.F. Balcan, and H.~Lin,
  editors, \emph{Advances in Neural Information Processing Systems}, volume~33,
  pages 18518--18529. Curran Associates, Inc., 2020.
\newblock URL
  \url{https://proceedings.neurips.cc/paper_files/paper/2020/file/d77c703536718b95308130ff2e5cf9ee-Paper.pdf}.

\bibitem[Yao et~al.(2021{\natexlab{b}})Yao, Dong, Zheng, Gholami, Yu, Tan,
  Wang, Huang, Wang, Mahoney, and Keutzer]{hawqv3}
Zhewei Yao, Zhen Dong, Zhangcheng Zheng, Amir Gholami, Jiali Yu, Eric Tan,
  Leyuan Wang, Qijing Huang, Yida Wang, Michael Mahoney, and Kurt Keutzer.
\newblock Hawq-v3: Dyadic neural network quantization.
\newblock In Marina Meila and Tong Zhang, editors, \emph{Proceedings of the
  38th International Conference on Machine Learning}, volume 139 of
  \emph{Proceedings of Machine Learning Research}, pages 11875--11886. PMLR,
  18--24 Jul 2021{\natexlab{b}}.
\newblock URL \url{https://proceedings.mlr.press/v139/yao21a.html}.

\bibitem[Hutchinson(1989)]{hutchinson1989stochastic}
Michael~F Hutchinson.
\newblock A stochastic estimator of the trace of the influence matrix for
  laplacian smoothing splines.
\newblock \emph{Communications in Statistics-Simulation and Computation},
  18\penalty0 (3):\penalty0 1059--1076, 1989.

\bibitem[Yao et~al.(2020{\natexlab{b}})Yao, Gholami, Keutzer, and
  Mahoney]{yao2020pyhessian}
Zhewei Yao, Amir Gholami, Kurt Keutzer, and Michael~W Mahoney.
\newblock Pyhessian: Neural networks through the lens of the hessian.
\newblock In \emph{2020 IEEE international conference on big data (Big data)},
  pages 581--590. IEEE, 2020{\natexlab{b}}.

\bibitem[Wu et~al.(2020)Wu, Xia, and Wang]{l2_flatness}
Dongxian Wu, Shu-Tao Xia, and Yisen Wang.
\newblock Adversarial weight perturbation helps robust generalization.
\newblock In H.~Larochelle, M.~Ranzato, R.~Hadsell, M.F. Balcan, and H.~Lin,
  editors, \emph{Advances in Neural Information Processing Systems}, volume~33,
  pages 2958--2969. Curran Associates, Inc., 2020.
\newblock URL
  \url{https://proceedings.neurips.cc/paper_files/paper/2020/file/1ef91c212e30e14bf125e9374262401f-Paper.pdf}.

\bibitem[LeCun et~al.(2010)LeCun, Cortes, and Burges]{lecun2010mnist}
Yann LeCun, Corinna Cortes, and CJ~Burges.
\newblock Mnist handwritten digit database.
\newblock \emph{ATT Labs [Online]. Available:
  http://yann.lecun.com/exdb/mnist}, 2, 2010.

\bibitem[Krizhevsky(2009)]{Krizhevsky09learningmultiple}
Alex Krizhevsky.
\newblock Learning multiple layers of features from tiny images.
\newblock Technical report, 2009.

\bibitem[Russakovsky et~al.(2015)Russakovsky, Deng, Su, Krause, Satheesh, Ma,
  Huang, Karpathy, Khosla, Bernstein, Berg, and Fei-Fei]{ILSVRC15}
Olga Russakovsky, Jia Deng, Hao Su, Jonathan Krause, Sanjeev Satheesh, Sean Ma,
  Zhiheng Huang, Andrej Karpathy, Aditya Khosla, Michael Bernstein,
  Alexander~C. Berg, and Li~Fei-Fei.
\newblock {ImageNet Large Scale Visual Recognition Challenge}.
\newblock \emph{International Journal of Computer Vision (IJCV)}, 115\penalty0
  (3):\penalty0 211--252, 2015.
\newblock \doi{10.1007/s11263-015-0816-y}.

\bibitem[Lecun et~al.(1998)Lecun, Bottou, Bengio, and Haffner]{lenet}
Y.~Lecun, L.~Bottou, Y.~Bengio, and P.~Haffner.
\newblock Gradient-based learning applied to document recognition.
\newblock \emph{Proceedings of the IEEE}, 86\penalty0 (11):\penalty0
  2278--2324, 1998.
\newblock \doi{10.1109/5.726791}.

\bibitem[He et~al.(2016)He, Zhang, Ren, and Sun]{he2016deep}
Kaiming He, Xiangyu Zhang, Shaoqing Ren, and Jian Sun.
\newblock Deep residual learning for image recognition.
\newblock In \emph{Proceedings of the IEEE conference on computer vision and
  pattern recognition}, pages 770--778, 2016.

\bibitem[Touvron et~al.(2021)Touvron, Cord, Douze, Massa, Sablayrolles, and
  J{\'e}gou]{touvron2021training}
Hugo Touvron, Matthieu Cord, Matthijs Douze, Francisco Massa, Alexandre
  Sablayrolles, and Herv{\'e} J{\'e}gou.
\newblock Training data-efficient image transformers \& distillation through
  attention.
\newblock In \emph{International conference on machine learning}, pages
  10347--10357. PMLR, 2021.

\bibitem[Jattke et~al.(2022)Jattke, Van Der~Veen, Frigo, Gunter, and
  Razavi]{jattke2022blacksmith}
Patrick Jattke, Victor Van Der~Veen, Pietro Frigo, Stijn Gunter, and Kaveh
  Razavi.
\newblock Blacksmith: Scalable rowhammering in the frequency domain.
\newblock In \emph{2022 IEEE Symposium on Security and Privacy (SP)}, pages
  716--734. IEEE, 2022.

\bibitem[Saileshwar et~al.(2022)Saileshwar, Wang, Qureshi, and
  Nair]{saileshwar2022randomized}
Gururaj Saileshwar, Bolin Wang, Moinuddin Qureshi, and Prashant~J Nair.
\newblock Randomized row-swap: Mitigating row hammer by breaking spatial
  correlation between aggressor and victim rows.
\newblock In \emph{Proceedings of the 27th ACM International Conference on
  Architectural Support for Programming Languages and Operating Systems}, pages
  1056--1069, 2022.

\bibitem[Han et~al.(2015)Han, Pool, Tran, and Dally]{pruning_first}
Song Han, Jeff Pool, John Tran, and William Dally.
\newblock Learning both weights and connections for efficient neural network.
\newblock In C.~Cortes, N.~Lawrence, D.~Lee, M.~Sugiyama, and R.~Garnett,
  editors, \emph{Advances in Neural Information Processing Systems}, volume~28.
  Curran Associates, Inc., 2015.
\newblock URL
  \url{https://proceedings.neurips.cc/paper_files/paper/2015/file/ae0eb3eed39d2bcef4622b2499a05fe6-Paper.pdf}.

\bibitem[Fiesler et~al.(1990)Fiesler, Choudry, and
  Caulfield]{fiesler1990weight}
Emile Fiesler, Amar Choudry, and H~John Caulfield.
\newblock Weight discretization paradigm for optical neural networks.
\newblock In \emph{Optical interconnections and networks}, volume 1281, pages
  164--173. SPIE, 1990.

\bibitem[LeCun et~al.(1989)LeCun, Denker, and Solla]{lecun1989optimal}
Yann LeCun, John Denker, and Sara Solla.
\newblock Optimal brain damage.
\newblock \emph{Advances in neural information processing systems}, 2, 1989.

\bibitem[Liu et~al.(2017)Liu, Li, Shen, Huang, Yan, and Zhang]{pruning_second}
Zhuang Liu, Jianguo Li, Zhiqiang Shen, Gao Huang, Shoumeng Yan, and Changshui
  Zhang.
\newblock Learning efficient convolutional networks through network slimming.
\newblock In \emph{Proceedings of the IEEE international conference on computer
  vision}, pages 2736--2744, 2017.

\end{thebibliography}
}

\newpage
\appendix
\onecolumn
\newpage
\appendix

\section{Detailed Experimental Setup}
\label{appendix:experimental-setup}

Here we describe our experimental setup in detail. All experiments use Python v3.11.4\footnote{Python: \href{https://www.python.org}{https://www.python.org}} with Pytorch v2.1.0\footnote{PyTorch: \href{https://pytorch.org/}{https://pytorch.org/}} and CUDA v12.1\footnote{CUDA: \href{https://developer.nvidia.com/cuda-downloads}{https://developer.nvidia.com/cuda-downloads}} 
for GPU acceleration.
We run our experiments on two systems:
(1) a node with a 48-core Intel Xeon Processor, 
768GB of memory, and 8 NVIDIA A40 GPUs. 
(2) a node with a 56-core Intel Xeon Processor, and 8 Nvidia Tesla H100 GPUs.
We achieve a substantial speed-up in running our evaluation script 
by utilizing the parameter-level parallelism on the two systems.

We use the following hyper-parameters to train/fine-tune our models.

\textbf{MNIST.} We use a network architecture (Base) and LeNet in prior work~\citep{hong2019terminal}. For regular training, we used an SGD optimizer with a learning rate of 0.1 (adjusting by 0.25 every 10 epochs), batch size of 64, and 0.8 momentum. We train our models for 40 epochs. %
To train the same network using our Hessian-aware training, we used $\lambda$ (line 16 of algorithm 1) value of 1 as per our findings in table 2. We use the RMSProp optimizer,  
keeping all the other hyper-parameters the same as the regular training.

\textbf{CIFAR-10.} We use ResNet18. For the regular training of this model, 
we use SGD, 0.02 learning rate, 32 batch-size, 0.9 momentum. We train our models for 90 epochs. 
We adjust the learning rate by 0.5 every 15 epochs. 
We use the RMSProp optimizer and $\lambda$ value of $10^{-2}$ to train the same model with our training method.

\textbf{ImageNet.} We take the ResNet50 architecture pretrained on ImageNet 
(available at Torchvision library\footnote{Pre-trained PyTorch models: \href{https://pytorch.org/vision/stable/models.html}{https://pytorch.org/vision/stable/models.html}}).
Instead of retraining the ResNet50 from scratch, 
we fine-tune the model on the same ImageNet dataset.
In fine-tuning, computing the Hessian matrix has a high computational demand.
We thus leverage our previous observation and focus on the layers closer to the model output.
We only compute Hessian eigenvalues and trace on the last layer and fine-tune the entire model using our training method. The hyper-parameters have been kept as Torchvision's original training hyper-parameters \footnote{\href{https://github.com/pytorch/vision/tree/main/references/classification}{https://github.com/pytorch/vision/tree/main/references/classification}}), but using the RMSProp optimizer. For fine-tuning the Diet-tiny ViT model on ImageNet, we use similar technique for hessian and eigenvalue computation. We take the pre-trained model from HuggingFace (available at \footnote{DeiT-tiny: \href{https://huggingface.co/facebook/deit-tiny-patch16-224}{https://huggingface.co/facebook/deit-tiny-patch16-224}}) and fine-tune it using our approach. We adopt the original training setup from~\citep{touvron2021training}, that uses batch size of 32, learning rate 0.1 and reducing by 0.1 every 30 epoch, momentum of 0.9, weight decay $10^{-4}$ and 90 epochs training cycle except we use the RMSProp optimizer. We experimentally found $\lambda$ value of $10^{-3}$ to achieve better generalization for our ImageNet model.

\section{Distribution Plot Computed on ImageNet Models}
\label{appendix:imagenet_distribution}

\begin{figure}[ht]
\centering
\includegraphics[width=0.5\textwidth]{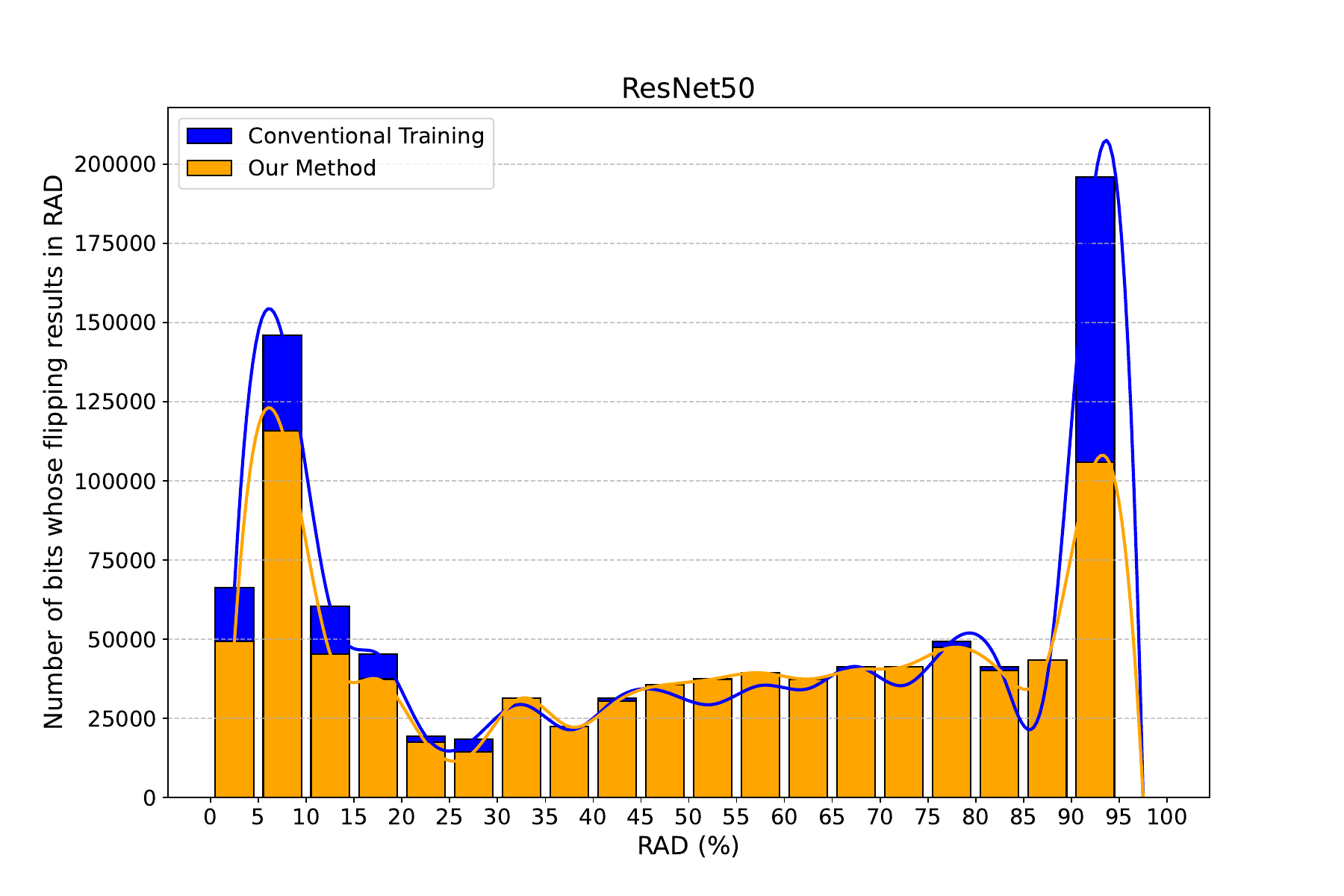}
\vspace{-2.em}
\caption{%
    \textbf{The distribution plot computed on ResNet50 in ImageNet.}
    Note that our fine-tuning only computes the Hessian trace from the last layer.}
\label{fig:imagenet_distribution}
\end{figure}
We show the distribution plot computed on the ImageNet models in figure~\ref{fig:imagenet_distribution}.
We observe that fine-tuning the pre-trained ResNet50 
achieves an enhanced resilience to bitwise errors in parameters.
It reduces the number of corruptions 
leading to an accuracy drop in the range between 0-30\%.
We also reduce the number of parameters 
whose bitwise error leads to an accuracy drop of over 90\% by half.
Our result on ImageNet is particularly interesting
because, even if we do not train our model with the Hessian trace 
computed on the entire layers, 
we can offer enhanced resilience to a DNN model.
While in MNIST and CIFAR-10 models, 
we see the number of parameters causing accuracy loss of 0--5\%, 
in our fine-tuned ImageNet model, 
we find a greater number of parameters
causing accuracy drops at 5--10\% bin.

\section{Visualizing Loss Landscapes of Layers with Residual Connections}
\label{appendix:loss-lanscape-resnets}

\begin{figure}[!h]
\centering
\begin{minipage}{0.48\linewidth}
    \centering
    \includegraphics[width=0.48\linewidth]{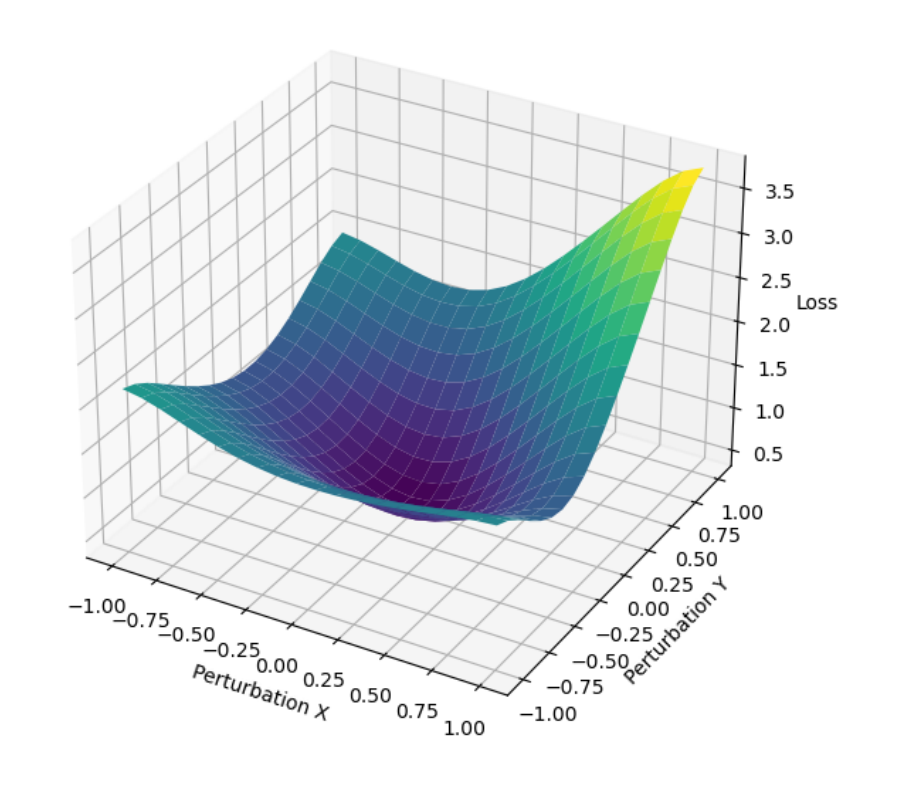}
    \includegraphics[width=0.48\linewidth]{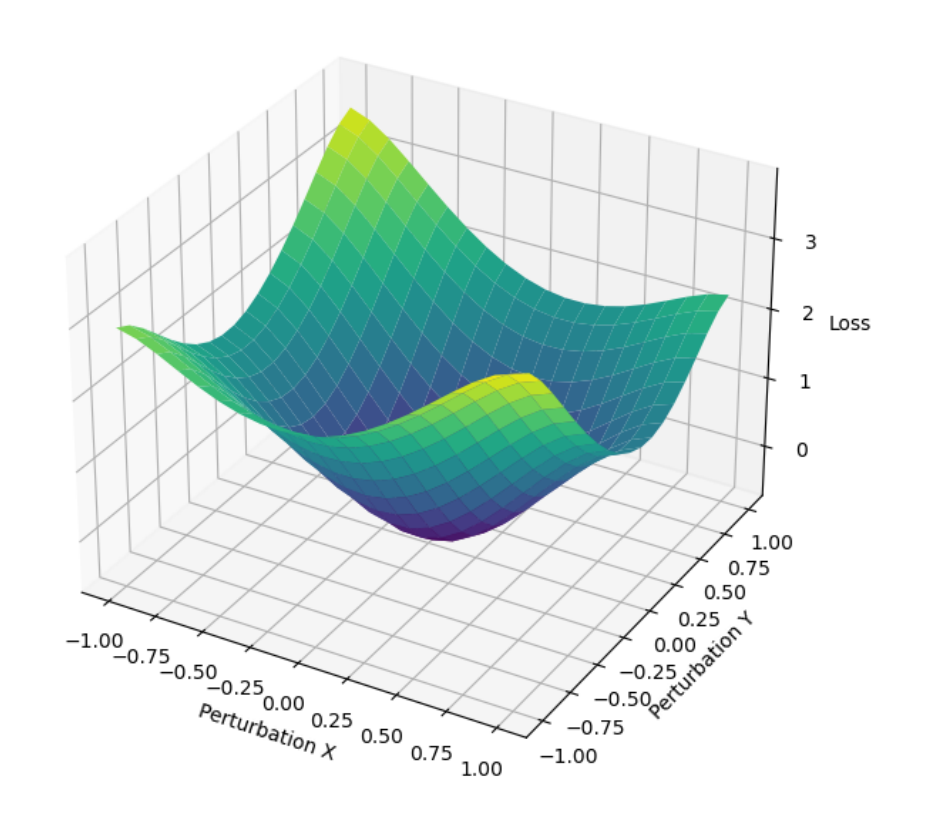}
\end{minipage}
\begin{minipage}{0.48\linewidth}
    \centering
    \includegraphics[width=0.48\linewidth]{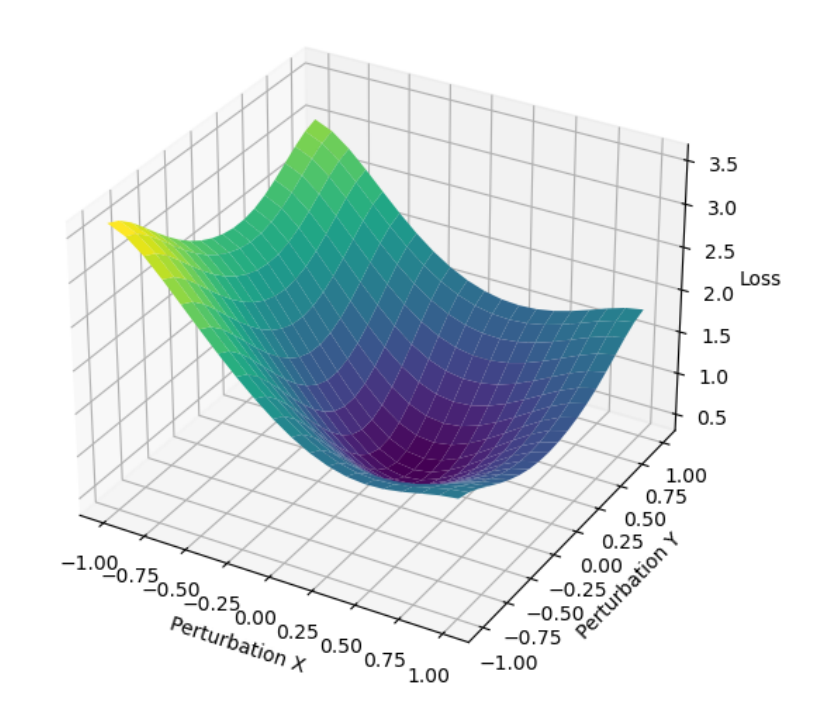}
    \includegraphics[width=0.48\linewidth]{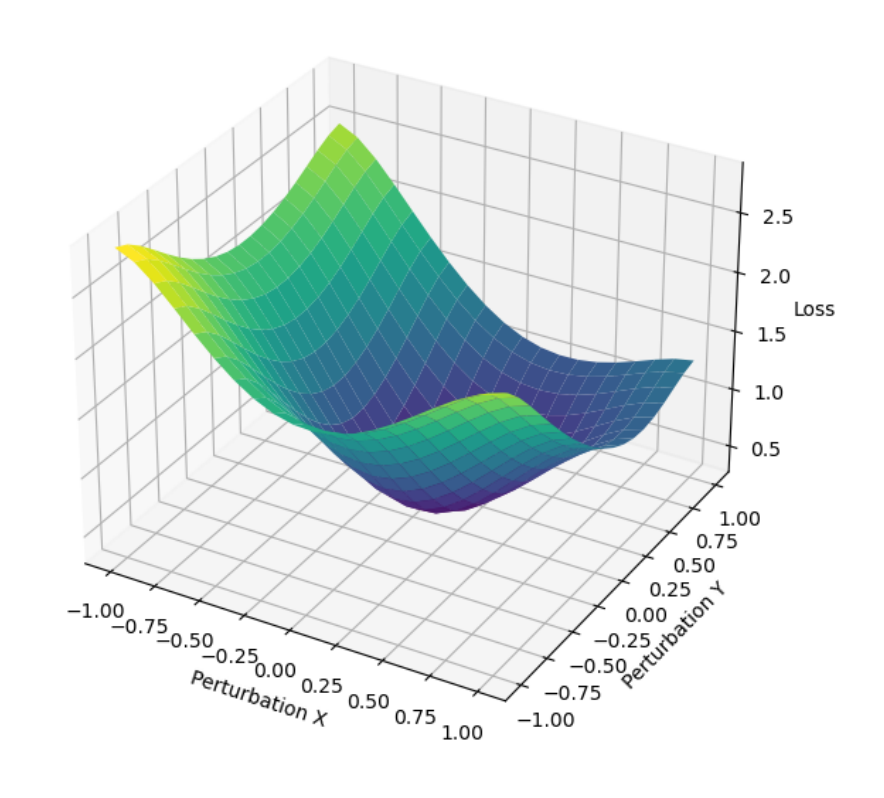}
\end{minipage}
\caption{%
\textbf{Comparing loss landscapes of the convolutional layers within a residual block.}
The left two are from the regularly-trained models, and the right ones are from those trained with our method.}
\label{fig:landscape-residual-plot}
\end{figure}

Prior work~\citep{li2018visualizing} has visually shown that 
convolutional layers with residual connections 
tend to have flatter loss surfaces.
In such layers, we hypothesize that
our approach is less effective in reducing the sensitivity.
Figure~\ref{fig:landscape-residual-plot} shows 
the loss landscapes from two convolutional layers
in ResNet18 models trained on CIFAR-10.
We observe that the loss landscapes 
visually look similar to each other,
implying that our approach was less effective
in reducing the Hessian trace of these layers.
This does not mean that these layers 
are particularly susceptible to bitwise errors in parameters.
On the other hands, 
these convolutional layers already have 
some resilience to bitwise errors in parameters.

\section{Numerical Perturbations Causing Accuracy Drop over 10\%}
\label{appendix:numerical_perturbation}

\begin{figure}[ht]
    \centering
    \begin{minipage}{0.33\textwidth}
        \includegraphics[width=\linewidth]{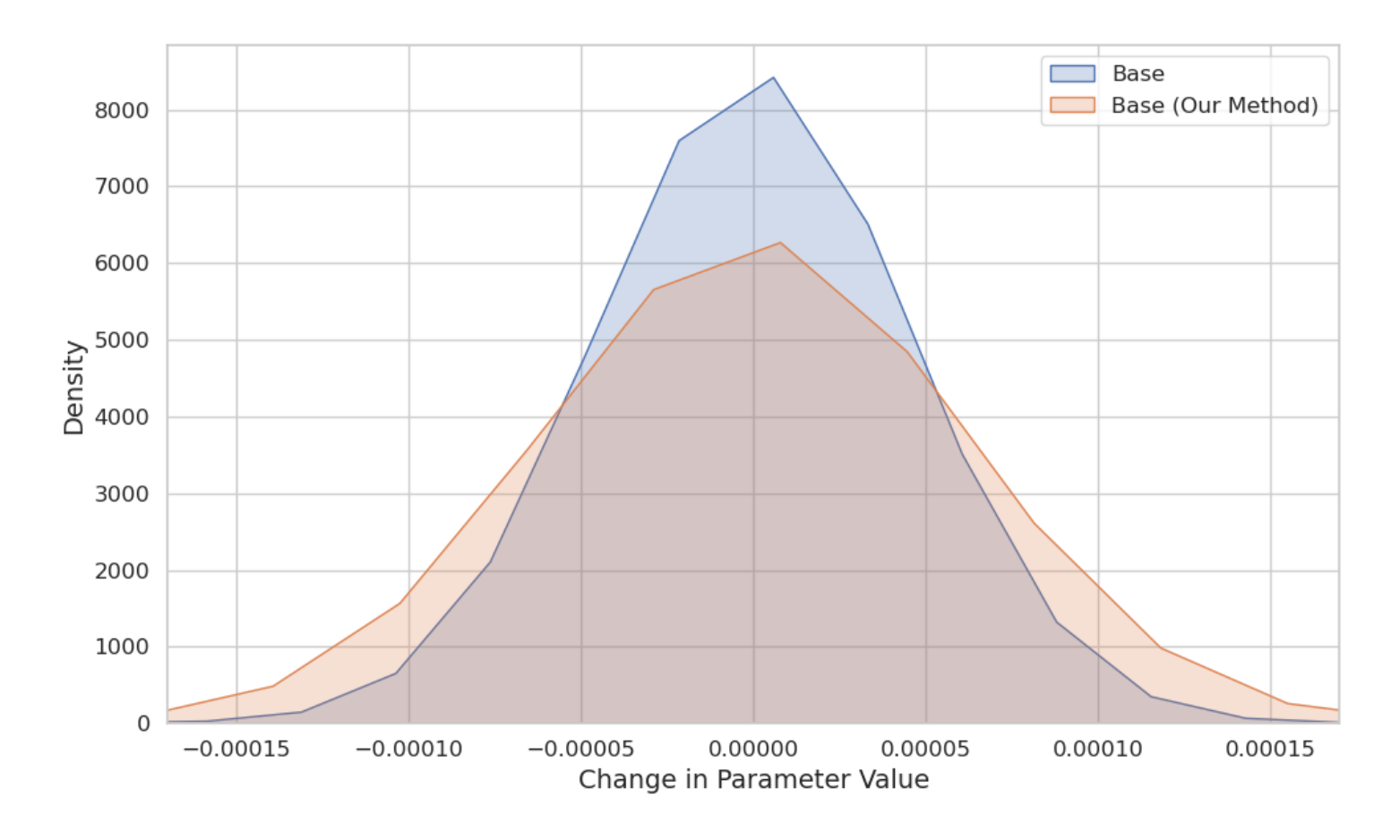}
    \end{minipage}%
    \hfill
    \begin{minipage}{0.33\textwidth}
        \includegraphics[width=\linewidth]{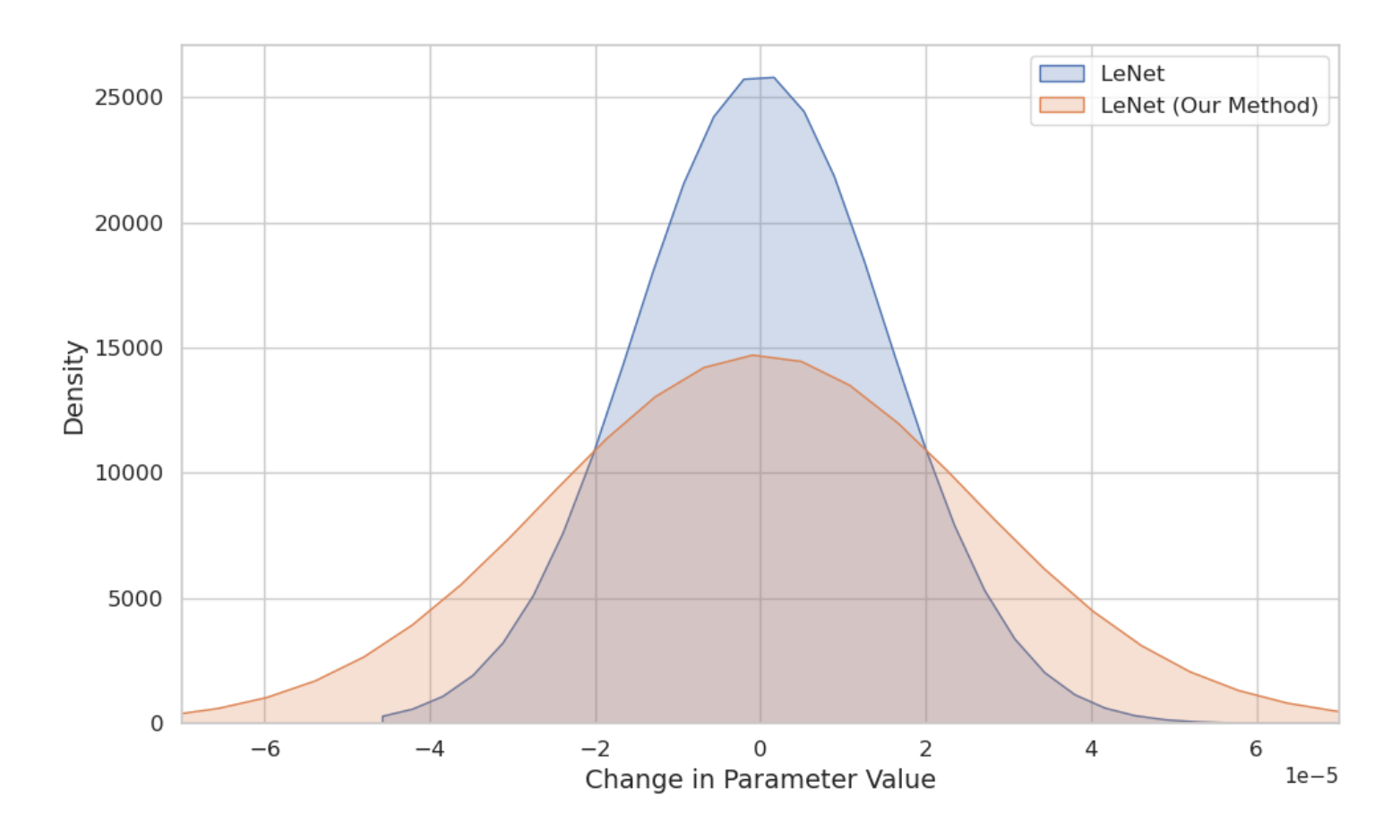}
    \end{minipage}%
    \hfill
    \begin{minipage}{0.33\textwidth}
        \includegraphics[width=\linewidth]{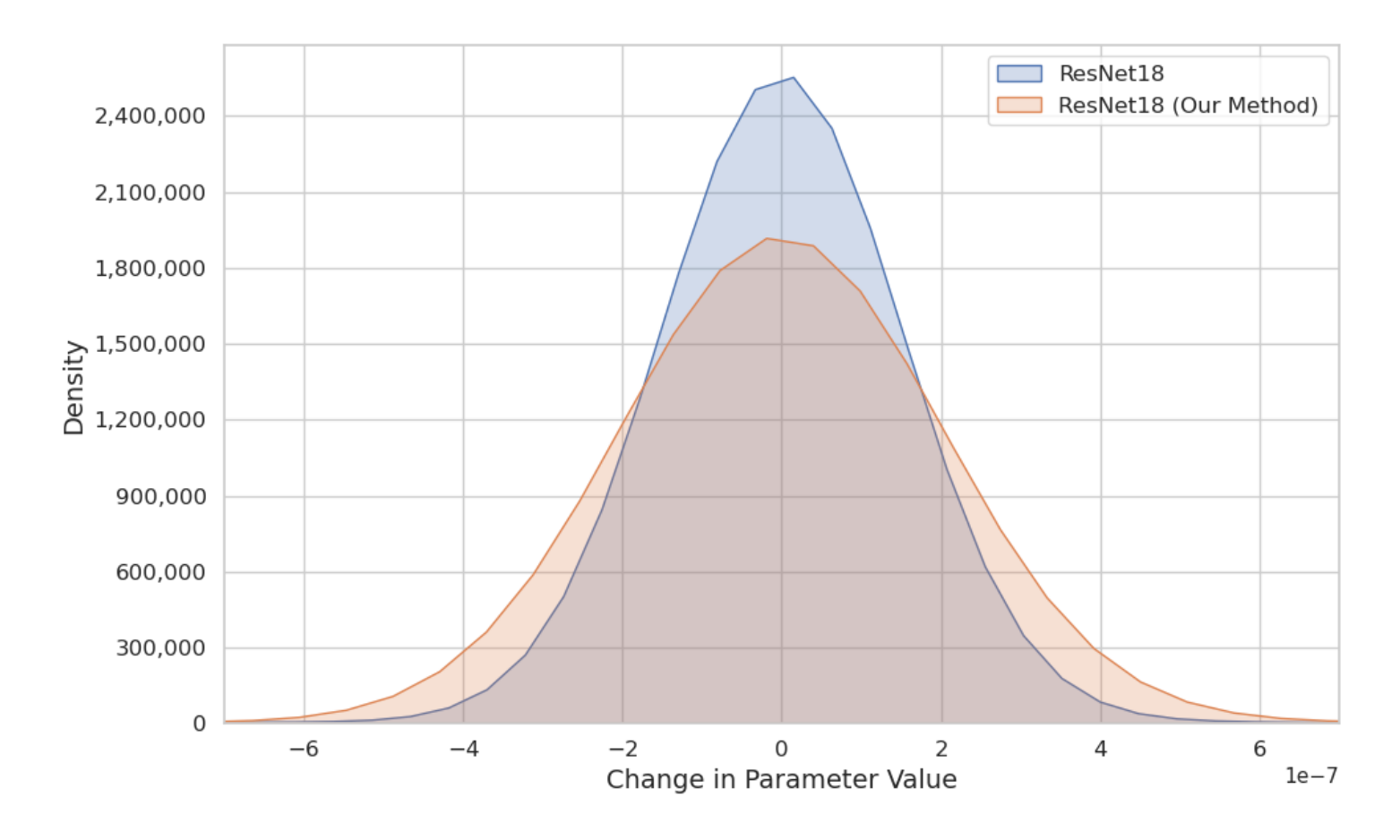}
    \end{minipage}
    \caption{\textbf{Comparison of numerical perturbations required to cause an accuracy drop over 10\%.}
    The left figure is computed on Base model, the middle one on LeNet, and the right one shows the result of ResNet18.}
    \label{fig:parameter-value-changes_resnet}
\end{figure}

We further analyze how resilient a model becomes 
to actual parameter value changes 
caused by single bitwise errors.
Using the parameter values before any corruption 
and after causing a single-bit error, 
we compute the changes in the numerical values on two models 
(one regularly-trained, and the other trained with our approach).
Figure~\ref{fig:parameter-value-changes_resnet} shows our results from the Base, LeNet and ResNet18 models.
We demonstrate that DNN models trained with our method
requires a greater numerical variations 
to cause a RAD drop over 10\%
than those trained using regular training methods.
Based on our observation that most single-bit errors cause a bit-flip in the most significant bit of the exponent (i.e., the \nth{31}-bit), the numerical variations required to cause a large performance loss go beyond the range
that floating-point representation in modern systems can hold.

\section{Overhead of Hessian Aware Training}
\label{appendix:overhead}
   
To measure the overhead we used the optimal hyper-parameter setup described in Appendix~\ref{appendix:experimental-setup} for all our models. We run training for 5 times, and report the per epoch training time. The measured values are presented in Table~\ref{tbl:performance}. Result shows that hessian aware training has a 4x-6x overhead for MNIST models. For the ResNet18 model trained on CIFAR-10, the overhead increases further due to ResNet18's larger architecture and higher number of parameters compared to smaller MNIST models. We employ layer sampling technique to reduce this overhead in our larger models. We calculate the Hessian eigenvalues and trace on the last layers and fine tune the model. Prior research~\citep{hong2019terminal} suggests that these final layers are the most susceptible against parameter corruption, making this a viable strategy for applying our method to large-scale models. Our result shows that adopting this method has only 1.18x computational overhead.

\begin{table}[ht]
\centering
\vspace{-0.2em}
\caption{\textbf{Comparing the training time of our method to baseline training in terms of runtime in PyTorch.} We report the per-epoch runtime (in seconds) for all our models trained across 3 datasets.}
\vspace{0.4em}
\adjustbox{max width=\linewidth}{
    \begin{tabular}{@{}cc|r|r@{}}
    \toprule
    \multirow{2}{*}{\textbf{Model}} &  \multirow{2}{*}{\textbf{Dataset}} & \multicolumn{2}{c}{\textbf{Training Time}} \\ \cmidrule{3-4} 
    
      &  & \textbf{Baseline} & \textbf{Our Method} \\ \midrule \midrule
    
    \textbf{Base}  & \multirow{2}{*}{MNIST} & 0.335 $\pm$ 0.002 & 1.362 $\pm$ 0.0085 \\ 

    \textbf{LeNet} &   &  0.432 $\pm$ 0.003 & 2.857 $\pm$ 0.0073 \\ 
    
    \textbf{ResNet18}  & CIFAR10 & 36.244 $\pm$ 0.607  & 341.58 $\pm$ 9.81 \\ 
    \textbf{ResNet50}  & ImageNet & 7275.6 $\pm$ 18.41 & 8647.2 $\pm$ 25.43 \\ \bottomrule
    \end{tabular}
}
\label{tbl:performance}
\vspace{-0.2em}
\end{table}

We conduct additional experiment on the layer-sampling technique for larger architecures like ResNet18 and ResNet50. Following the same overhaed measurement approach, We applied Hessian regularization incrementally to the layers closer to output. We start with only the last layer and extend it to the last 2, 3, and finally 4 layers of the model and compared the runtime with baseline training. Our results are presented in Table~\ref{tbl:overhead}. 

\begin{table}[ht]
\centering
\vspace{-0.2em}
\caption{\textbf{Comparing the training time of layer-sampling and baseline training in PyTorch.} We report the per-epoch runtime (in seconds).}
\vspace{0.4em}
\adjustbox{max width=\linewidth}{
    \begin{tabular}{@{}cc|c|c| c |c | c}
    \toprule
    \multirow{2}{*}{\textbf{Model}} &  \multirow{2}{*}{\textbf{Dataset}} & \multicolumn{2}{c}{\textbf{Training Time (in seconds)}} \\ \cmidrule{3-7} 
    
      &  & \textbf{Baseline} & \textbf{L1}  & \textbf{L2} & \textbf{L3} & \textbf{L4} \\ \midrule \midrule
    
    \textbf{ResNet18}  & CIFAR10 & 36.244 $\pm$ 0.607  & 37.77 $\pm$ 0.39 & 43.24 $\pm$ 0.28 & 57.63 $\pm$ 0.44 & 78.24 $\pm$ 1.13\\ 
    \textbf{ResNet50}  & ImageNet & 7275.6 $\pm$ 18.41 & 8647.2 $\pm$ 25.43 & 10134.7 $\pm$ 30.21 & 13289.5 $\pm$ 35.76 & 16547.8 $\pm$ 42.15\\ \bottomrule
    \end{tabular}
}
\label{tbl:overhead}
\vspace{-0.2em}
\end{table}

Results in Table~\ref{tbl:overhead} demonstrate that training overhead increases as we increase the “layers involved in Hessian eigenvalue and trace calculation.” However, using only the last 1 layer of the model, we can reduce the overhead to almost the same as baseline training, making our method efficient for very large models. We note that the increased computational time is not solely due to adopting our training method. The additional time is primarily attributed to the computation of the large hessian trace and eigenvalues, which is not fully optimized for use with popular deep learning frameworks such as PyTorch. Further optimization of our approach will be an interesting future work.

\section{Analysis of Corrupted Bit Position}
\label{appendix:bit-position}

\begin{figure}[ht]
\centering
\begin{minipage}{0.32\linewidth}
    \includegraphics[width=\linewidth]{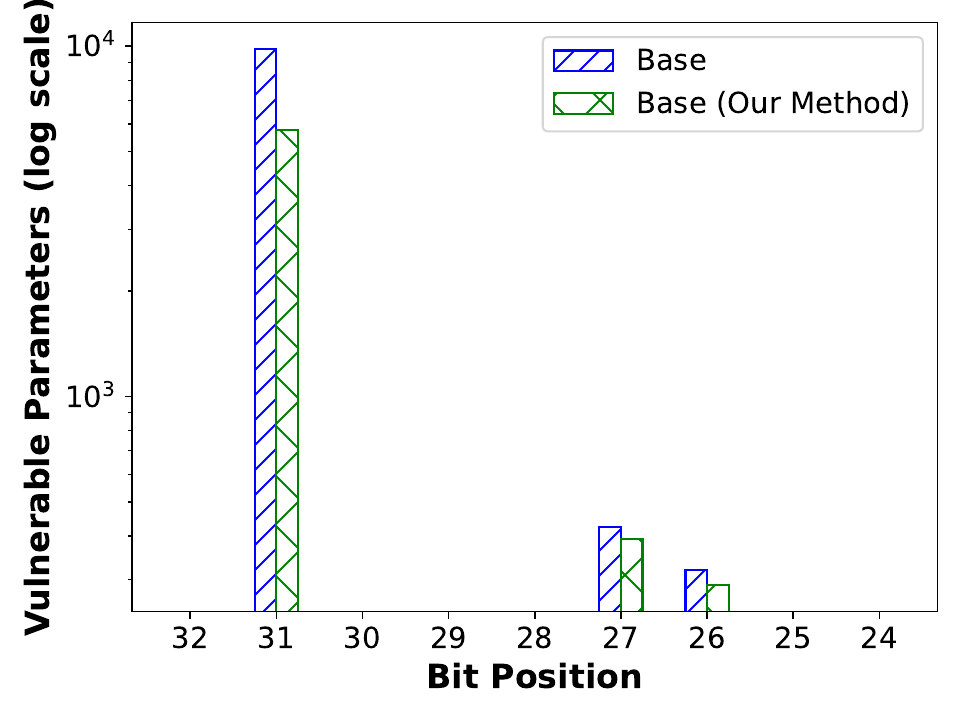}
\end{minipage}
\begin{minipage}{0.32\linewidth}
    \includegraphics[width=\linewidth]{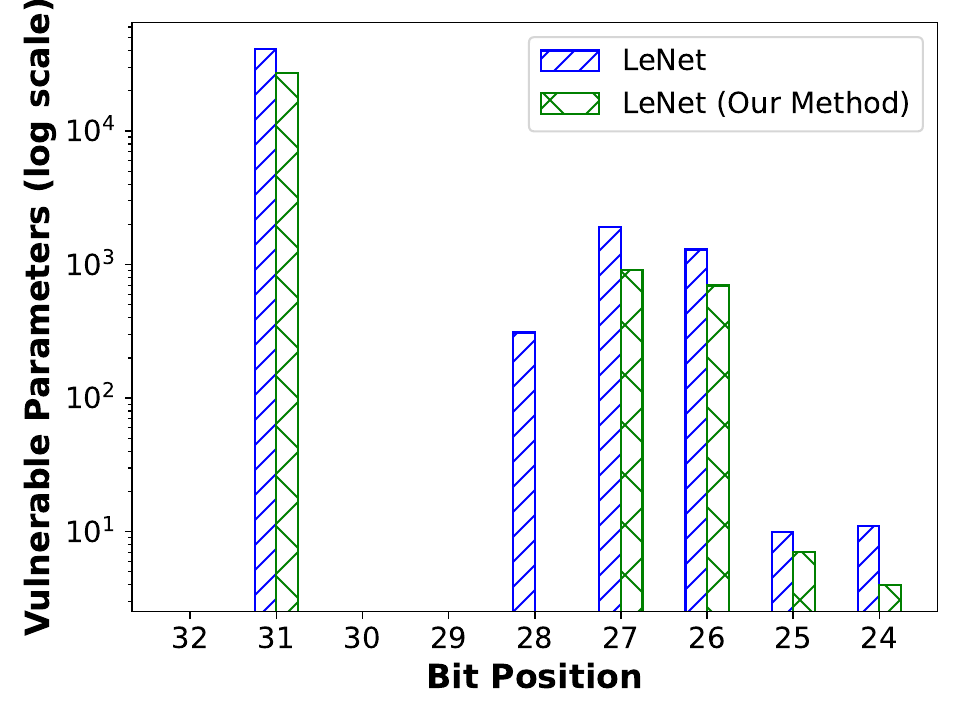}
\end{minipage}
\begin{minipage}{0.32\linewidth}
    \centering
    \includegraphics[width=\linewidth]{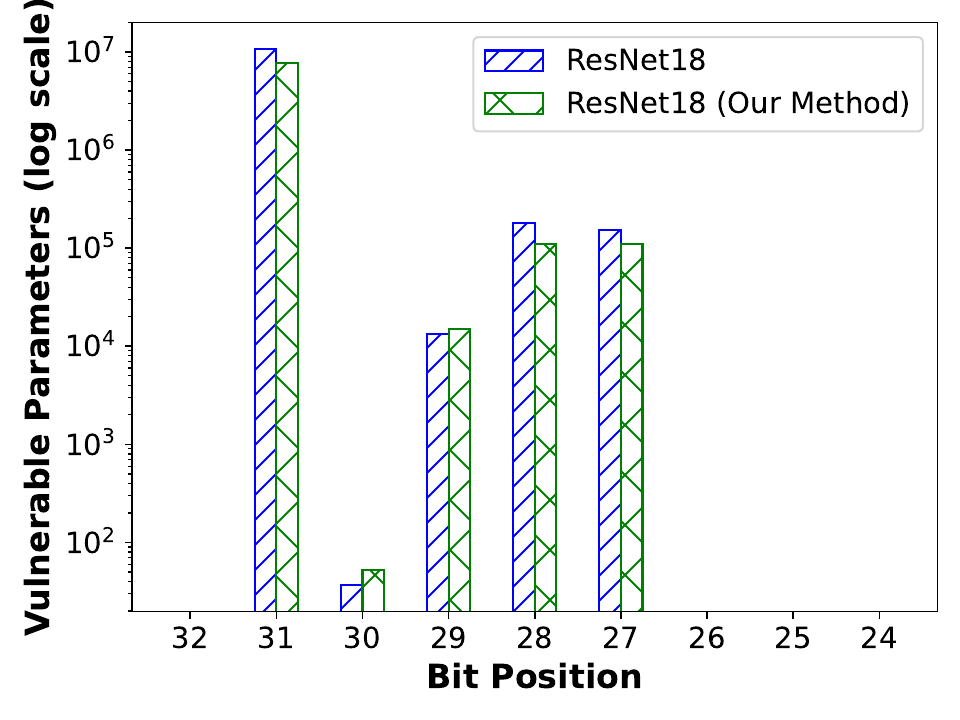}
\end{minipage}
\caption{%
\textbf{Comparison of the corrupted bit positions.} 
From left to right, we show the analysis result from Base (MNIST), LeNet (MNIST), and ResNet18 (CIFAR-10).
We only examine the sign bit and the exponent bits, as they change the numerical value of a parameter the most.}
\label{fig:bit_position}
\end{figure}

The IEEE 754 standard defines the representation of floating-point numbers in modern computer systems. 
In this format, a 32-bit number is represented with three fields: 
the 1-bit sign, the 8-bit exponent, and the 23-bit mantissa.
Similar to the prior work~\citep{hong2019terminal, rakin2019bit, deephammer}, 
we analyze the location of bitwise corporations 
that lead to an accuracy drop over 10\%. 
Figure~\ref{fig:bit_position} shows our analysis results.
We use a logarithmic scale in the y-axis.

In all the models, corruption of the $31^{st}$ bit mostly leads to an accuracy drop over 10\%.
These corruptions account for $\sim$93\% and $\sim$91.43\% 
in the Base and LeNel models, respectively.
We also observe a few bits in the $26^{th}$ and $27^{th}$ position 
for both Base and LeNet models 
and a small number of bits 
in the $28^{th}$ location for the LeNet model. 
A consistent trend is observed in the ResNet18 models in CIFAR-10, 
with the $31^{st}$ bit being identified as the most susceptible bit location.
However, in ResNet18, we identify a few bits positioned 
at the $30^{th}$ and $29^{th}$ location in the exponent.
In contrast to our observations from LeNet and ResNet18,
there are no susceptible corruptions in
the $30^{th}, 29^{th}, 28^{th}, 25^{th}$ and $24^{th}$ bit positions
in the Base model.

\section{Enhanced Model Resilience to Compression}
\label{appendix:pruning-quantization}

We examine the additional benefits of our approach 
beyond parameter resilience to bitwise errors.
We are particularly interested in testing 
whether models trained with our method 
can achieve improved performance under 
pruning~\citep{pruning_first} 
or quantization~\citep{fiesler1990weight}.
These techniques reduce the size of neural networks 
through parameter reduction or compression, 
introducing optimal parameter perturbations~\citep{lecun1989optimal}.
Although it is not the focus of our work, we study the effectiveness of our method in increasing the resilience of DNN models against these perturbations.

\begin{figure}[!ht]
\centering
\begin{minipage}{0.33\textwidth}
    \includegraphics[width=\linewidth]{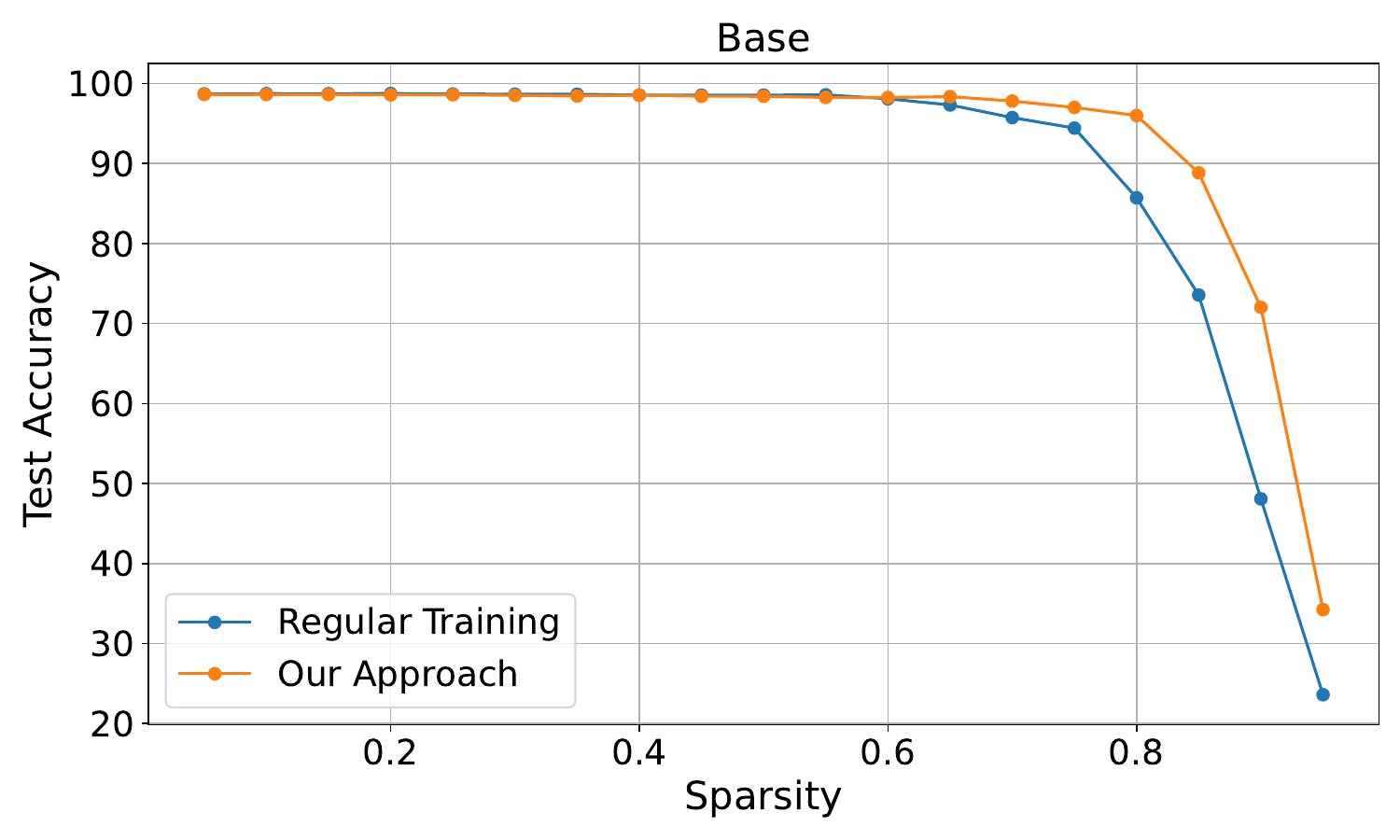}
\end{minipage}%
\hfill
\begin{minipage}{0.33\textwidth}
    \includegraphics[width=\linewidth]{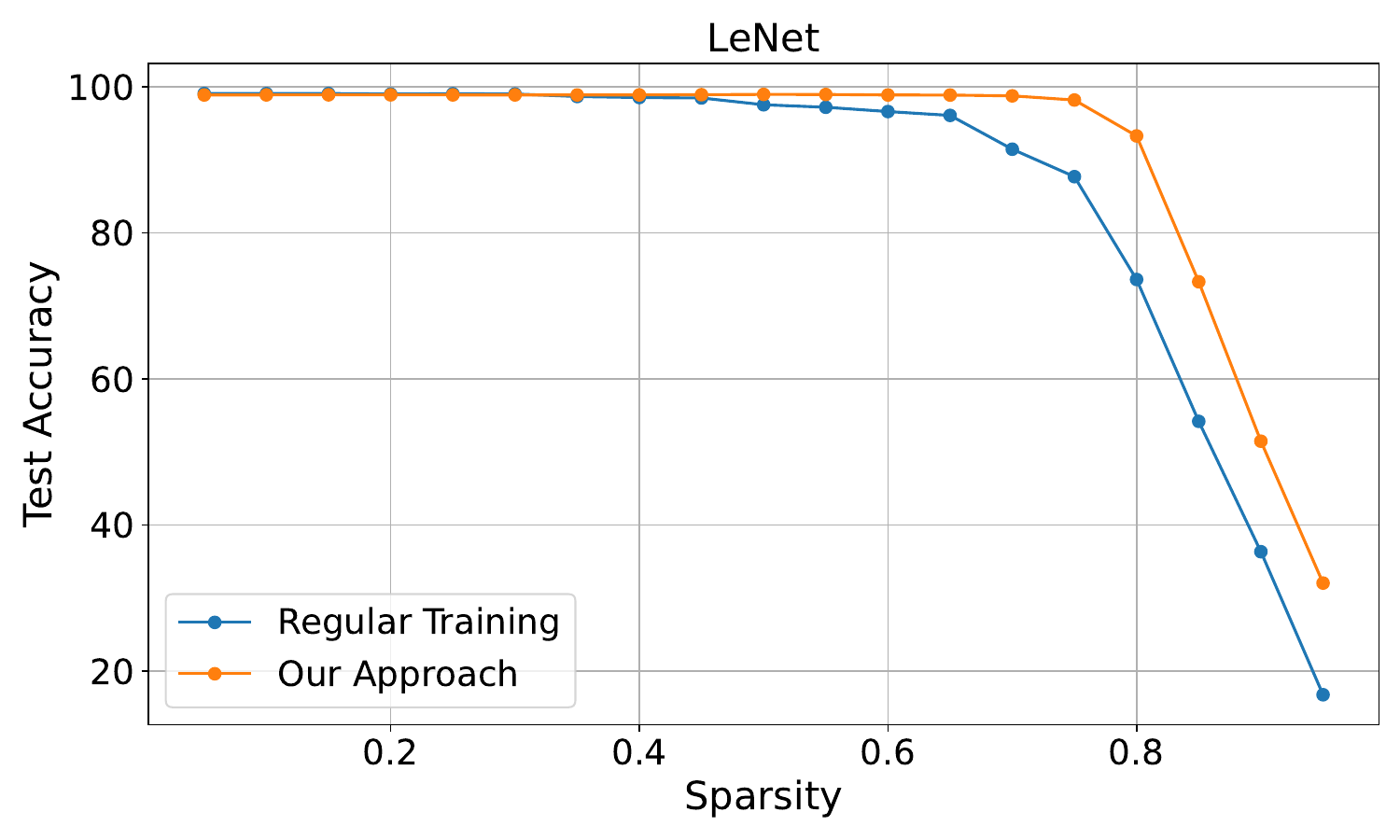}
\end{minipage}
\begin{minipage}{0.33\textwidth}
    \includegraphics[width=\linewidth]{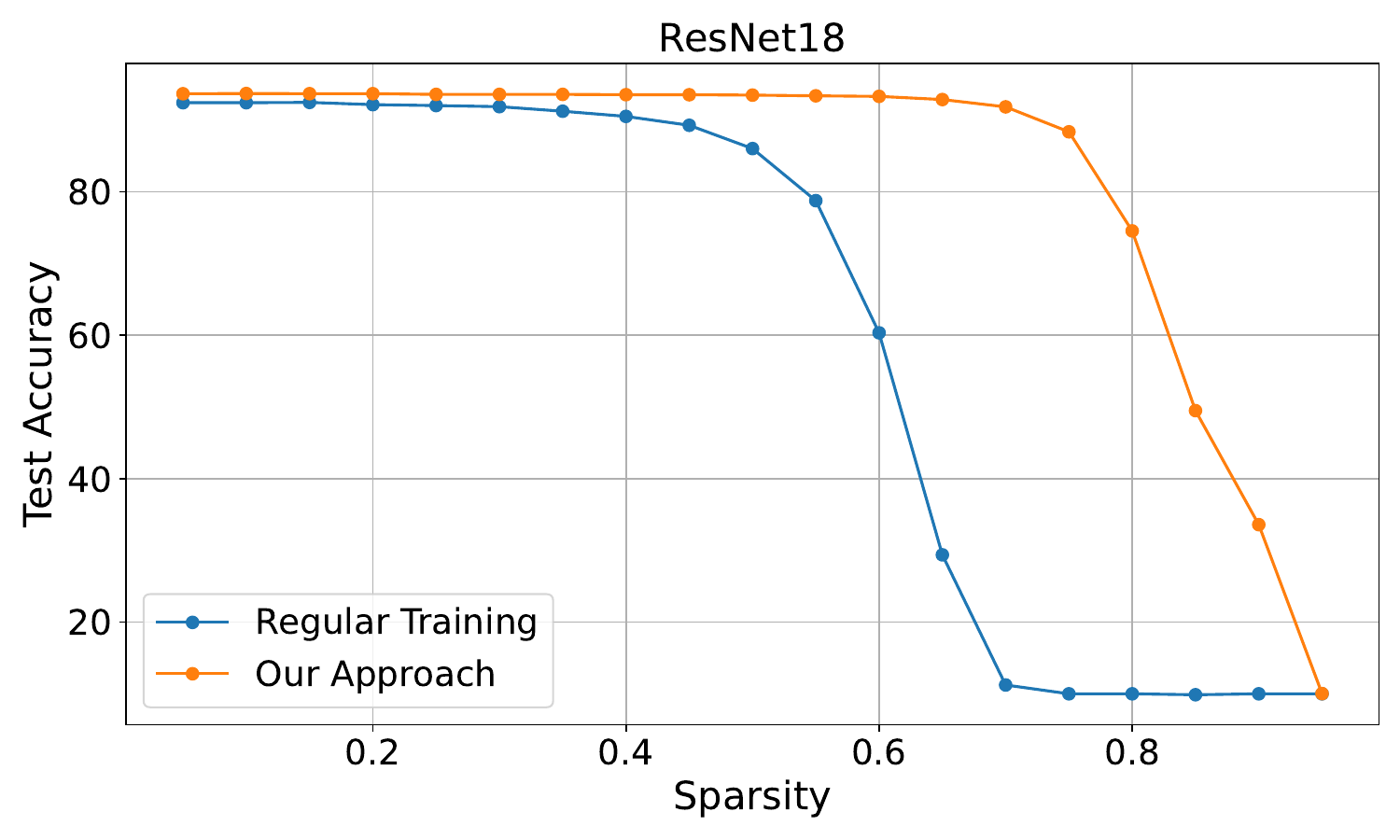}
\end{minipage}
\caption{\textbf{Comparison of model performance under various pruning ratios.}
The left and middle figure (Base and LeNet Model) is computed on the MNIST dataset, while the right one is from the ResNet18 CIFAR-10 models.}
\label{fig:pruning_comparison_base_lenet}
\end{figure}

\textbf{Pruning.}
In our evaluation, we employ global unstructured pruning~\citep{pruning_second}, 
which operates at the individual weight level.
This technique first computes an importance score 
for each weight and removes those with the lowest scores.
We apply this pruning technique with 
different sparsity levels ranging from 0--100\%.
Figure~\ref{fig:pruning_comparison_base_lenet} shows 
our pruning results on the Base and LeNet model on MNIST and ResNet18 models trained on CIFAR-10.
We demonstrate that DNN models trained with our method 
retain accuracy better than those trained using regular training methods.
Both MNIST models retains their original accuracy up to 65\% parameters pruned.
Beyond this point, as sparsity increases, 
we observe a steep decrease in accuracy.
The Base and LeNet models trained using our method
shows better accuracy than the regularly-trained models.  
Our approach surprisingly maintains accuracy further for ResNet18 model on CIFAR-10, up to 70\% pruning indicating enhanced parameter-level resilience to bitwise errors. At the same sparsity level, the model trained with the conventional approach 
completely loses accuracy (i.e., the accuracy dropping to $\sim$0\%).

\textbf{Quantization.} 
Table \ref{tab:quantization} summarizes our quantization results 
for 8-, 4-, and 2-bit quantization of 
the regularly-trained models and Hessian-aware trained models.
We employ layer-wise, symmetric quantization, 
which is the default in most deep learning frameworks.
Overall, the models trained with our approach achieve 
better test accuracy than the regularly trained models, an additional benefit that hessian-aware training offers.
Up to 4-bit quantization, both models retain 
the performance of their floating-point counterparts.
However, when we use 2-bit precision, 
the accuracy of all models decreases significantly.
Our models under 2-bit precision consistently achieve 
1.5--14\% better accuracy, indicating that 
these models have increased resilience to parameter value variations.
Based on our observation that fully-connected layers 
are less sensitive than convolutional layers (see the above analysis), 
we employ mixed-precision quantization with 
2-bit precision in fully-connected layers 
and 4-bit precision in convolutional layers.
We demonstrate that our models achieve an accuracy of 68.8--78.7\%, 
while the regularly-trained models achieve 48.9--68.2\% model accuracy.
\begin{table}[ht]
\centering
\vspace{-1.4em}
\caption{%
\textbf{Comparison of model performance under various quantization ratios.}
We compare the test accuracy of models after quantizing them with different bit-widths.
}
\label{tab:quantization}
\vspace{0.2em}
\adjustbox{max width=\linewidth}{
\begin{tabular}{ll|ccc|c}
\toprule
\multirow{2}{*}{\textbf{Dataset}} & \multirow{2}{*}{\textbf{Model}} & \multicolumn{4}{|c}{\textbf{Acc.}} \\ \cmidrule{3-6} 
 & & \textbf{8-bit} & \textbf{4-bit} & \textbf{2-bit} & \textbf{Mixed} \\ \midrule \midrule
\multirow{4.5}{*}{\textbf{MNIST}} & \textbf{Base} & 98.57 & 98.38 & 24.49 & 48.90 \\
                       & \textbf{Base (Ours)} & 98.73 & 98.70 & 38.72 & 68.84 \\ \cmidrule{2-6} 
                       & \textbf{LeNet} & 99.10 & 98.70 & 11.85 & 57.03 \\
                       & \textbf{LeNet (Ours)} & 98.90 & 97.37 & 24.78 & 73.90 \\ \midrule
\multirow{2}{*}{\textbf{CIFAR-10}} & \textbf{ResNet18} & 92.53 & 88.01 & 9.96 & 68.19 \\
                         & \textbf{ResNet18 (Ours)} & 92.36 & 90.26 & 10.28 & 78.69 \\ \bottomrule
\end{tabular}
}
\end{table}

\end{document}